\documentclass[iop,numberedappendix,useAMS,usenatbib]{emulateapj}
\usepackage{times}
\usepackage{apjfonts}
\usepackage{amssymb}
\usepackage{epsfig}

\def\HI{{H{\small I }}}

\begin{document}

\shorttitle{Effects Of The Ionosphere On Ground-Based Global 21 CM Signal Detection}
\shortauthors{Datta et al.}

\title{Effects Of The Ionosphere On Ground-Based Detection Of The Global 21 CM Signal From The Cosmic Dawn And The Dark Ages}

\author{Abhirup Datta\altaffilmark{1,2}, Richard Bradley\altaffilmark{3}, Jack O. Burns\altaffilmark{1}, Geraint Harker\altaffilmark{4}, Attila Komjathy \altaffilmark{5} and T. Joseph W. Lazio\altaffilmark{5}}

\altaffiltext{1}{Center for Astrophysics and Space Astronomy, Department of Astrophysical and Planetary Science, University of Colorado, Boulder, C0 80309,USA}
\altaffiltext{2}{Center of Astronomy, Indian Institute of Technology Indore, Khandwa Road, Simrol Campus, Indore-452020, India}
\altaffiltext{3}{National Radio Astronomy Observatory, 520 Edgemont Road, Charlottesville, VA 22903, USA}
\altaffiltext{4}{Department of Physics and Astronomy, University College London, Gower Street, London WC1E 6BT}
\altaffiltext{5}{Jet Propulsion Laboratory, California Institute of Technology, 4800 Oak Grove Drive, Pasadena, CA 91109, USA}

\email{Abhirup.Datta@colorado.edu}

\begin{abstract}
Detection of the global \HI 21 cm signal from the Cosmic Dawn and the Epoch of Reionization is the key science driver for several ongoing ground-based and future ground- /space- based experiments. The crucial spectral features in the global 21 cm signal (turning points) occur at low radio frequencies $\lesssim 100$~MHz. In addition to the human-generated RFI (Radio Frequency Interference), Earth's ionosphere drastically corrupts low-frequency radio observations from the ground. In this paper, we examine the effects of time-varying ionospheric refraction, absorption and thermal emission at these low radio frequencies and their combined effect on any ground-based global 21 cm experiment. It should be noted that this is the first study of the effect of a dynamic ionosphere on global 21 cm experiments. The fluctuations in the ionosphere are influenced by solar activity with flicker noise characteristics. The same characteristics are reflected in the ionospheric corruption to any radio signal passing through the ionosphere. As a result, any ground based observations of the faint global 21 cm signal are corrupted by flicker noise (or ``$1/f$'' noise, where ``$f$'' is the dynamical frequency) which scales as $\nu^{-2}$ (where $\nu$ is the frequency of radio observation) in the presence of a bright galactic foreground ($\propto \nu^{-s}$, where $s$ is the radio spectral index). Hence, the calibration of the ionosphere for any such experiment is critical. Any attempt to calibrate the ionospheric effects will be subject to the inaccuracies in the current ionospheric measurements using GPS (Global Positioning System) ionospheric measurements, riometer measurements, ionospheric soundings, etc. Even considering an optimistic improvement in the accuracy of GPS-TEC (Total Electron Content) measurements, we conclude that the detection of the global 21 cm signal below 100 MHz is best done from above the Earth's atmosphere in orbit of the Moon. 
\end{abstract}


\section{Introduction}
Detection of the highly redshifted $\lambda$21 cm ``spin-flip'' transition \citep{field58} of the neutral hydrogen (\HI) against the Cosmic Microwave Background (CMB) is considered as a promising probe for the cosmic Dark Ages ($z\gtrsim 30$), the Cosmic Dawn ($30\gtrsim z \gtrsim 15$), and the Epoch of Reionization ($15 \gtrsim z \gtrsim 6$). Studying the early universe ($z \gtrsim 6$) through the redshifted 21 cm signal will allow us to understand the nature of the first stars, galaxies and black holes \citep{madau97, furlanetto06, pritchard12}. 
\begin{figure*}[t!]
\centering
\epsfig{file=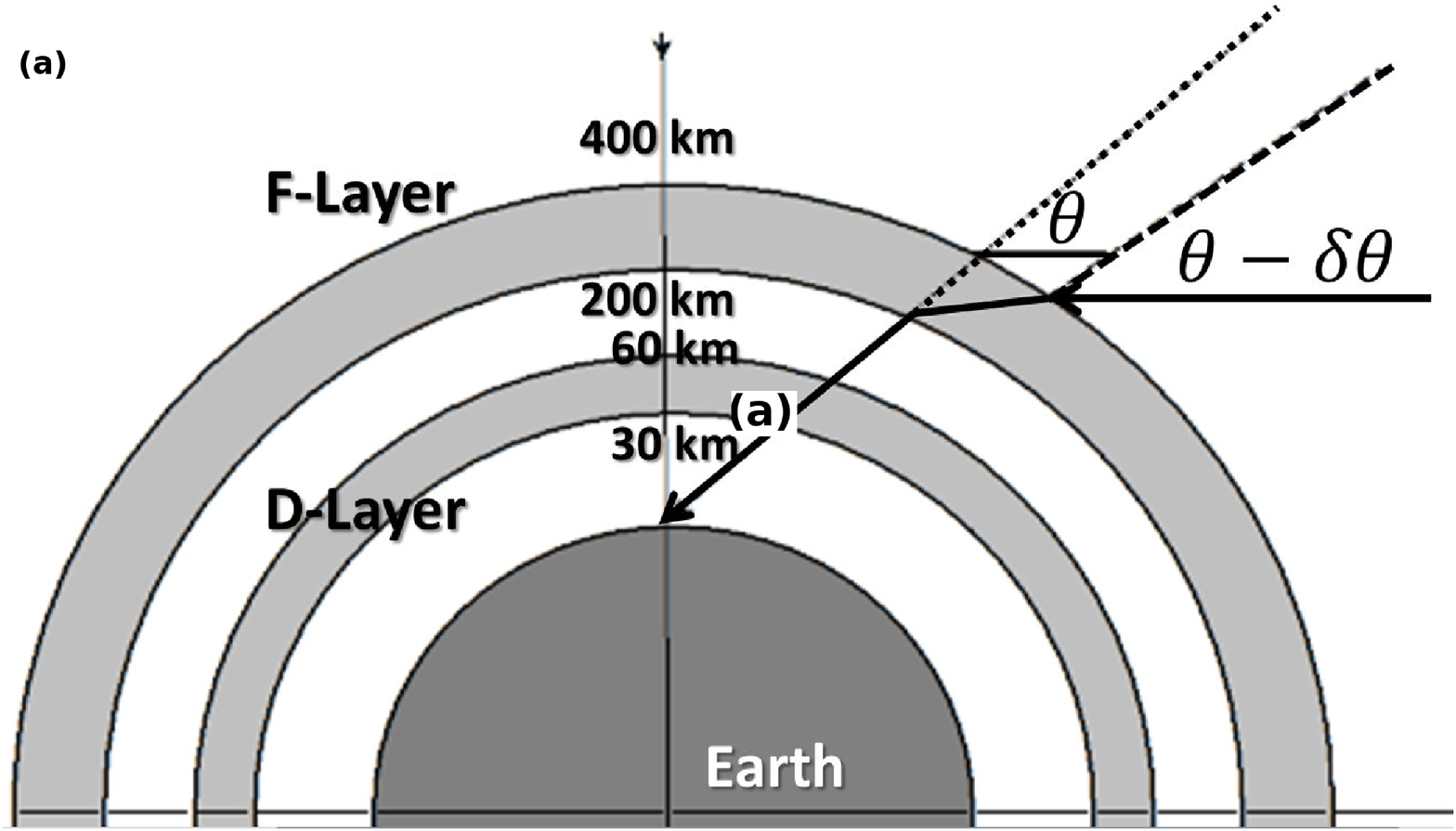,height=2.4truein,width=3.2truein}
\epsfig{file=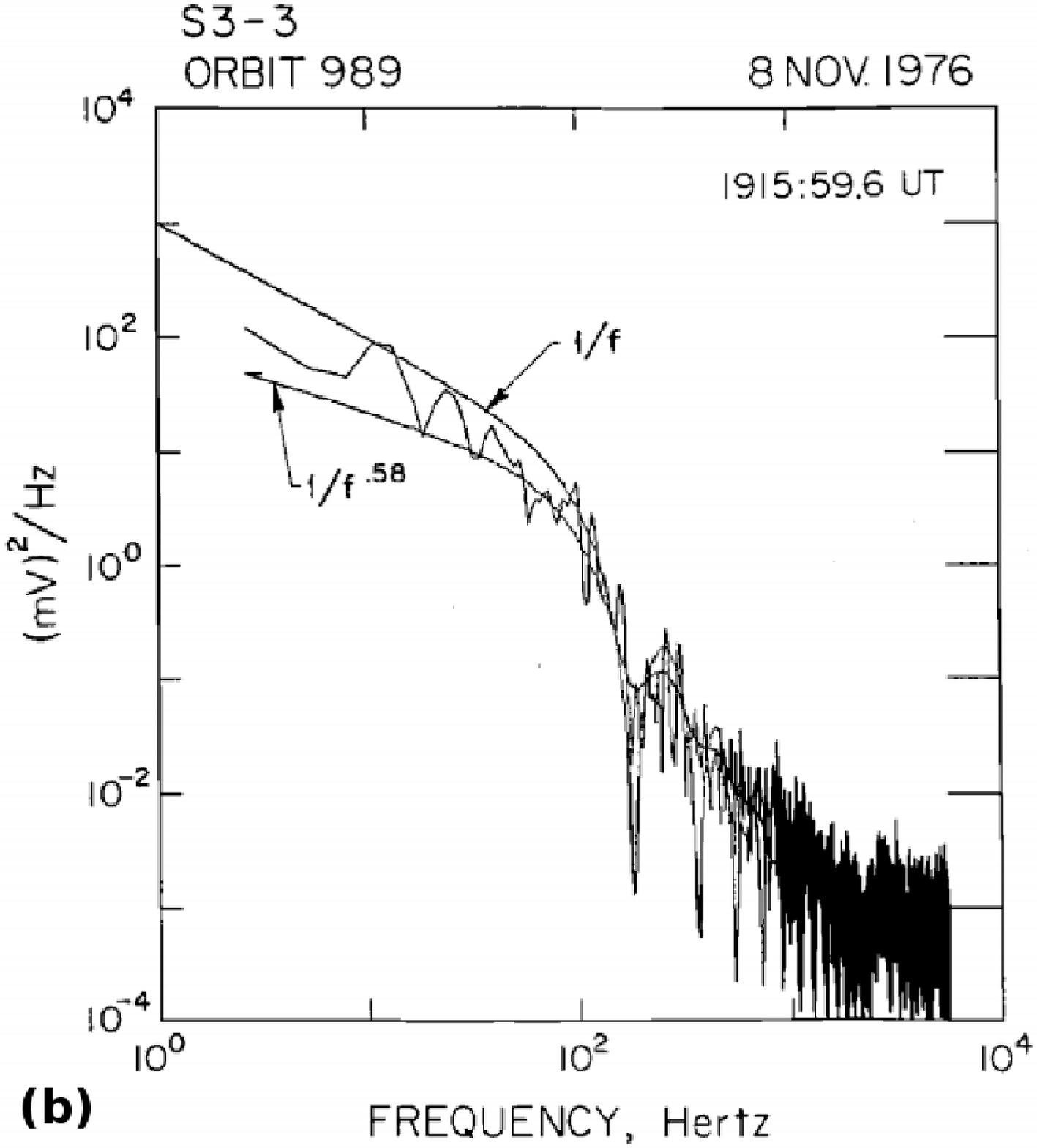,height=2.4truein,width=2.6truein}
\caption{{\bf (a)} Schematic representation (not to scale) of the Earth's ionosphere showing the 'F' and the 'D' layers which are responsible for refraction and attenuation/emission respectively \citep{harish14}. Also shown in this Figure, the effect of ionospheric refraction on incident rays. {\bf (b)} Electric field power spectrum from the S33-satellite from \citet{temerin89} (This figure is same as Figure 5 in \citet{temerin89} and re-used here with proper permission from the publishers.) The power spectrum shows a $1/f^\alpha$ characteristics and looks similar to the power spectrum of the electron density fluctuations in the ionosphere over GreenBank, WV (Figure~\ref{fig:TEC_var_hist_GB}(b))}
\label{fig:iono}
\end{figure*}

There are two different approaches to observe this signal: (a) using large interferometric arrays at these low radio frequencies to produce statistical power spectra of the \HI 21 cm fluctuations \citep{pober13,paciga13,hazelton13,harker10} and possibly using images of the \HI 21 cm fluctuations \citep{zaroubi12}, or (b) using a single antenna at low radio frequencies to detect the ``all-sky'' averaged \HI 21 cm signal as a function of redshift \citep{shaver99}. In this paper, we will concentrate only on the second approach. 

\begin{figure*}[t!]
\centering
\epsfig{file=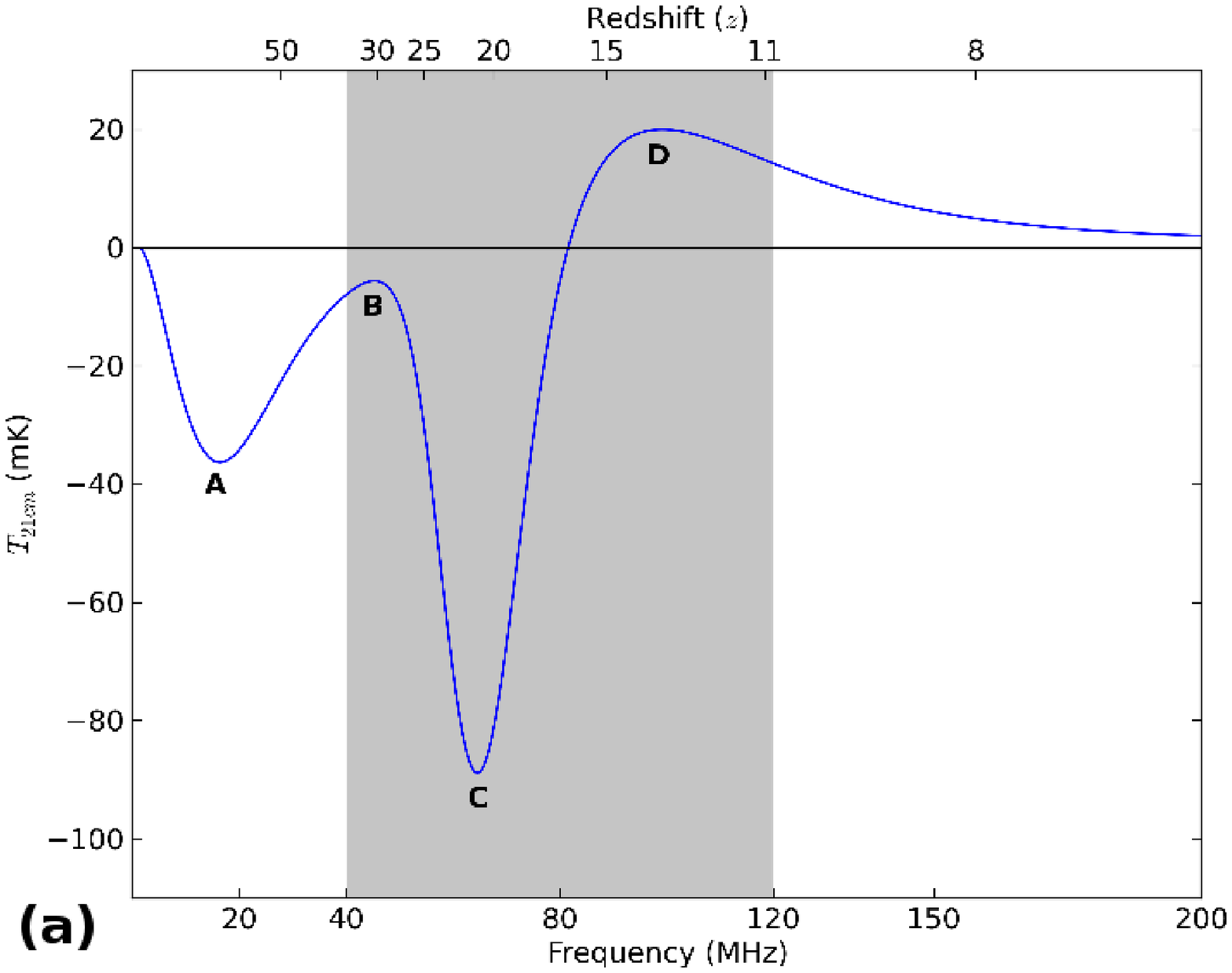,height=2.4truein}
\epsfig{file=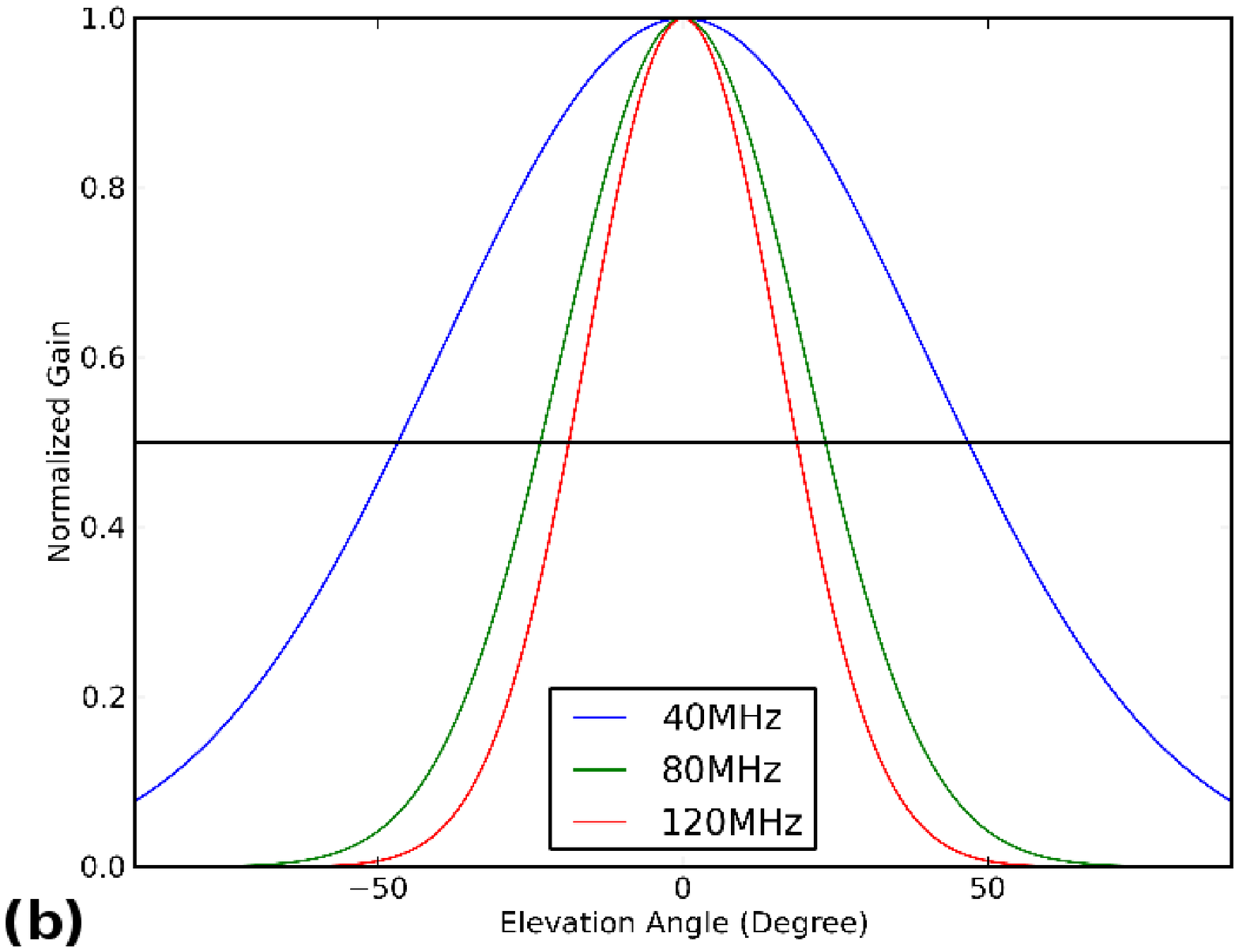,height=2.4truein}
\caption{{\bf (a)} The model 21 cm ``all-sky'' averaged signal showing the turning points `B', `C' and `D' (reference model of \citet{mirocha14}). {\bf (b)} Symmetric Gaussian primary beam for the fiducial instrument at 40, 80 and 120 MHz.}
\label{fig:sig}
\end{figure*}

Several ground-based experiments are underway to detect the global 21 cm signal from the Epoch of Reionization and Cosmic Dawn, such as Experiment to Detect the Global EoR Signature -- EDGES \citep{bowman08, bowman10}, Shaped Antenna measurement of the background RAdio Spectrum  -- SARAS \citep{patra13}, Large Aperture Experiment to Detect the Dark Age -- LEDA \citep{bernardi14}, ``Sonda Cosmol\'ogica de las Islas para la Detecci\'on de Hidr\'ogeno Neutro''--SCI-HI \citep{voytek14} and Broadband Instrument for Global Hydrogen Reionization Signal -- BIGHORNS \citep{sokolowski15a}. Although this second approach is conceptually simpler than the radio interferometric approach, detection of this faint cosmological \HI signal ($\sim10-100$~mK) with a single antenna needs to achieve dynamic ranges of $\sim 10^4 - 10^6$ in the presence of strong Galactic and extragalactic foregrounds ($\gtrsim 10^3 -10^4$~K). In addition, ground-based experiments will be affected by human-generated RFI (Radio Frequency Interferences), such as the FM-band (87.5-110 MHz) which falls in the middle of this observed spectrum (Figure~\ref{fig:sig}), and the effects resulting from the signals having passed through the Earth's ionosphere. 

The ionosphere is a part of the upper atmosphere stretching from $\sim 50-600$ km above the Earth's surface. The electron densities in the ionosphere change significantly due to the effect of solar activity \citep{evans68,ratcliffe72,davies90}. The presence of the Earth's ionosphere results in three effects relevant for the detection of the redshifted HI 21cm signal.  The ionosphere refracts all trans-ionospheric signals including the Galactic and extragalactic foregrounds, causes attenuation to any trans-ionospheric signal \citep{evans68,davies90} and also produces thermal emission \citep{pawsey51,steiger61}. Moreover, these effects are intrinsically time variable due to the solar forcing of the ionosphere \citep{evans68,ratcliffe72,davies90}. Since these ionospheric effects scale as $\nu^{-2}$, where $\nu$ is the frequency of observations, these effects are expected to be more pronounced for the detection of the global 21 cm signal from the Cosmic Dawn and the Dark Ages ($z \gtrsim 15$) than the same from the Epoch of Reionization($15 \gtrsim z \gtrsim 6$).

\citet{rogers11m} and \citet{harish14} have previously considered a subset of these effects and their implications for the detection of the global 21 cm signal using ground-based experiments. \citet{rogers11m} outlined the effects of attenuation and emission due to a static ionosphere on a global 21 cm signal observation above 100 MHz. \citet{harish14} studied the effects of refraction and absorption due to a static ionosphere on ground-based global 21 cm experiments between 30 and 100 MHz. Using a simple ionospheric model, \citet{harish14} showed that the additional foregrounds introduced due to Earth's ionosphere are 2-3 orders of magnitude higher than the expected 21 cm signal. In a more recent study, \citet{rogers15} detected the effects of a dynamic ionosphere on EDGES observations in Western Australia. They derived the differential opacity and electron temperature in the ionosphere.

In this paper, we investigate the challenges for global 21 cm signal detection below 100 MHz from the ground in the presence of a dynamic (time-variable) ionosphere with the goal of assessing the extent to which a ground-based experiment is even feasible. In Section 2 of this paper, we review Earth's ionosphere and its interaction with solar activity. In Section 3, we discuss the parameters involved in the simulations performed in this paper. Section 4 discusses the effect of Earth's ionosphere on global signal observations through refraction, absorption and emission. In Section 5, we discuss the effect of a typical night-time ionosphere on global 21 cm signal detection as well as the effect of the uncertainties in the ionospheric measurements on ionospheric calibration.    

\section{Earth's Ionosphere}
\label{sec:iono}

Earth's ionosphere can be divided into several layers (see Figure~\ref{fig:iono}(a)): D-layer (60-90 km), composite F-layer (160-600 km) and E-layer (which lies between the D and F layers). Earth's ionosphere is naturally influenced by solar activity.  

The Sun radiates in a wide range of the electromagnetic spectrum, ranging from radio wavelengths to infrared, visible, ultraviolet, x-ray and beyond. The solar ultraviolet light and soft/hard X-rays interact with Earth's upper atmosphere and its constituents through photo-ionization processes \citep{evans68,ratcliffe72,davies90}. This interaction causes the formation of an ionized layer called the ionosphere. The ionization in the ionosphere is mostly due to solar UV radiation and partly due to cosmic rays. The UV radiation of the Sun ionizes the F-layer of the ionosphere, while the soft X-rays from the Sun ionizes the E-layer. The D-layer is ionized by the hard X-ray component of the solar radiation. In addition, solar flares and solar wind cause changes in the ionization level in various layers of the ionosphere \citep{davies90}.  

 Based on the nature of the solar disturbances, the electron densities and temperature in the ionosphere change significantly \citep{evans68,ratcliffe72,davies90}. The solar activity follows variabilities at different temporal scales. The variability in the dynamical system of the ionosphere is a direct consequence of the forcing action by the solar radiation. Thus, the ionosphere will also reflect the same scales of solar temporal variability \citep{ozgucc08, liu11} through ionospheric turbulence, scintillation, etc. It is well known that the various solar activities such as solar radio bursts and even sun-spot index display ``$1/f$'' characteristics (see Appendix A) as a function of time \citep{ryabov97, planat01, polygiannakis03}. Even during times of relatively little solar activity, the variability of the solar forcing produces variations in the ionospheric electron density and temperature that display characteristics 1/f or flicker noise. As a result, the variations in the electron density and temperature also display ``1/f'' (or flicker) noise characteristics \citep{surkov08,zhou11,roux11} reflecting the effects of solar activity \citep{elkins69,yeh82,temerin89, truhlik15}. The electron density in the various layers of the ionosphere has a well-understood, quadratic dependence on the plasma frequency or $\nu_p$ (defined later in equation~\ref{eq:plasma}), and long duration radiosonde measurements taken from Slough, England from 1932-1963 show $\nu_p$ variability on time scales ranging from hours to years \citep{davies90}. Such low frequency fluctuations exhibiting dynamical behavior on logarithmic scales is the hallmark of ``$1/f$'' distributions \citep{barnes66, williams04, schmid08}. A flicker noise does not have a well-defined mean over long times and it moves further away from the initial value as time progresses (e.g. \citet{press78}). Also, a flicker noise does not reduce as $\propto 1/\sqrt{N_{\textnormal{samples}}}$ or $1/\sqrt{\delta t}$ (where $N_{\textnormal{samples}}$ is the number of samples corresponding to a integration time of $\delta t$), unlike Gaussian noise. In Appendix~\ref{sec:pink}, we discuss the basic theory of a ``1/f'' process or flicker noise relevant to our analysis of the ionosphere.  

Figure~\ref{fig:iono}(b) shows the power spectrum of electric field fluctuations in the ionosphere taken between 0 and 18.6 kHz at a sample rate of 0.37s by the S33 polar orbiting satellite  \citep{temerin89}. The resultant electric field power spectrum from this observations of ionospheric turbulence clearly shows a $1/f^{0.6}$ trend.

The F-layer also consists of F-1 and F-2 layers, extending up to 1000 km from the Earth's surface. However, for our simulations we only consider a single layer for F extending between 200 and 400 km which contributes the most significantly to the total electron content of the F-layer \citep{bilitza03,bilitza15,harish14}. The F-layer is characterized by low atmospheric gas density and high electron density. Thus the collision rate in the F-layer is low. On the other hand, the D-layer has high atomic gas density and low electron density. Hence, the collision rate in the D-layer is high. The attenuation of radio waves in the ionosphere is caused by collisions of the electrons with ions and neutral particles \citep{evans68}. Thus the D-layer mainly contributes to the attenuation of radio signals passing through the ionosphere. Since the extent of the F-layer is larger than the D-layer, any trans-ionospheric signal suffers multipath propagation while traveling through the F-layer. Hence, the F-layer contributes mainly to the ionospheric refraction. In our simulations, we consider (a) ionospheric refraction due to the F-layer, (b) attenuation/emission due to the D-layer \citep{hsieh66a}. The existence of the E-layer is strongly dependent on the solar activity but it is also likely to be present even during the night-time. In this paper, we only consider the effects of the F and D layers of the ionosphere as they dominate the effects of the refraction and absorption/emission respectively.

\section{Simulations}
In order to understand the effect of the Earth's ionosphere on the Global 21 cm experiments from the ground, we included a model 21 cm signal, a simple primary beam model of a fiducial telescope and a model foreground sky. Here, we describe these simulation parameters.

\subsection{Global 21 CM Signal}
The redshifted, sky-averaged (i.e. 'global') 21 cm signal ($T_{21cm}$), expressed as a differential brightness temperature relative to the CMB (Cosmic Microwave Background), depends on the mean neutral hydrogen fraction ($\overline{x_{\HI}}$) and is given by \citep{furlanetto06}:
\begin{equation}
T_{21cm} = 27\overline{x_{\HI}}\left(\frac{T_s - T_\gamma}{T_s}\right) \left(\frac{1+z}{10}\right)^{1/2} ~~\textnormal{mK}  
\label{eq:sig}
\end{equation}
where $T_s$ is the 21 cm spin temperature and $T_{\gamma}$ is the CMB temperature. Figure~\ref{fig:sig}(a) shows a model 21 cm signal (reference model of \citet{mirocha14}) that will be used in the simulations for this paper. This model 21 cm signal is qualitatively similar to realizations appearing in recent literature and should be treated as just a representative model. We follow the nomenclature of \citet{pritchard10} and refer to the ``critical'' points in the global 21 cm spectrum as Turning Points A,B,C and D (Figure~\ref{fig:sig}(a)). The Turning Points are useful as diagnostics of the global 21 cm signal \citep{harker12}, and also as model-independent tracers of IGM properties \citep{mirocha14}.

Since the ionospheric effects scale as $\nu^{-2}$ where $\nu$ is the frequency of observations, the effect on detection of Turning Point A is expected to be much worse than that on B. Hence, in this paper, we limit the lowest frequency of interest to 40 MHz which excludes Turning Point A. Also, at higher frequencies ($\gtrsim 100$~MHz) the ionospheric effects are expected to be less. Hence, we have restricted the highest frequency of interest to 120 MHz which still includes Turning Point D (according to the model shown in Figure~\ref{fig:sig}(a)). Therefore, in this paper, we limit our frequency band of interest between 40-120 MHz which includes Turning Points B,C and D. 

\subsection{Instrumental Beam Model}
\label{sec:beam}
In order to carry out the simulations, we have assumed an ideal instrument with symmetric Gaussian beam pattern (Figure~\ref{fig:sig}(b)). The half power beam-width (HPBW) of the primary beam at 75 MHz is $\sim 60^o$ and scales as $\nu^{-1}$. Hence, the field-of-view of the observations increases as the frequency of observations decreases.

This ideal beam pattern is chosen here to demonstrate the effect of ionosphere. If more realistic beam shapes are considered, the effects will be worse than shown in this paper.

\subsection{Foregrounds}
The most important foreground for global 21 cm experiments is the diffuse emission from the Galaxy and other galaxies. Galactic synchrotron emission contributes $\sim 70\%$ of the total foreground while the extragalactic emission contributes $\sim 27\%$ of the total foregrounds \citep{jelic08}. These two components dominate the system temperature of any global 21 cm experiments at these low radio frequencies. The large primary beam (see Section~\ref{sec:beam}) will average over a wide section of the sky. In this paper, we have only included the diffuse emission in the foreground. Any inclusion of the extragalactic point sources will only increase the total sky temperature as measured by the instrument which will further increase the additional sky temperature due to ionospheric effects (see Section~\ref{sec:effects}).

The diffuse foreground spectra have been derived following the treatment in \citet{harker12}. The primary beam model for the fiducial instrument has been convolved with the Global Sky Map of \citet{oliveira08} to derive a foreground spectrum given by: 
\begin{eqnarray}
T_{FG}(\nu,\Theta_0,\Phi_0) & = & \int_0^{2\pi} d\Phi \int_0^{\pi/2} d\Theta B(\nu,\Theta-\Theta_0,\Phi-\Phi_0) \nonumber \\
& & T_{GSM}(\nu,\Theta-\Theta_0,\Phi-\Phi_0) \sin{\Theta}
\label{eq:conv}
\end{eqnarray}
where $T_{FG}(\nu,\Theta_0,\Phi_0)$ is the convolved spectrum for one pointing $(\Theta_0,\Phi_0)$ in the Global Sky Map ($T_{GSM}(\nu,\Theta,\Phi)$) and $B(\nu,\Theta-\Theta_0,\Phi-\Phi_0)$ denotes the original primary beam power pattern which peaks at $(\Theta_0,\Phi_0)$ (Figure~\ref{fig:sig}(b)). It should be noted that the Galactic foreground has an angular dependence which results in variation in the sky spectrum when convolved with different widths of the model primary beam. This is essential to consider when computing the effect of the ionospheric refraction on the increase in the sky temperature as seen by a ground-based telescope (see Section~\ref{sec:refrac}).

Combining equations~\ref{eq:sig} and \ref{eq:conv}, we obtain the resultant sky temperature as:
\begin{equation}
T_{sky}(\nu)=T_{FG}(\nu) +T_{21cm}(\nu)
\label{eq:sky}
\end{equation}

The thermal noise on the simulated observations is derived from the radiometer equation:
\begin{equation}
\sigma (\nu) = \frac{T_{sys}(\nu)}{\sqrt{\delta \nu * \delta t}}
\label{eq:therm_noise}
\end{equation}
where $\delta \nu = 0.5$~MHz is the channel bandwidth and $\delta t$ is the time over which the given spectrum is averaged over. Thermal noise values will be used in our simulations in Section~\ref{sec:measure} to estimate the additional noise introduced by the ionosphere for any global 21 cm signal experiments.  It should be noted here that at these low radio frequencies system temperature of radiometer is dominated by the brightness temperature of the sky, i.e. $T_{sys} \approx T_{sky}$.

\section{Effect of The Ionosphere On Global Signal Detection}
\label{sec:effects}

The intensity of any electromagnetic wave passing through medium like the ionosphere, which is generally optically thin, obeys the radiative transfer equation \citep{sysbook01}. The corresponding brightness temperature of the trans-ionospheric radio signal can be written as:
\begin{eqnarray}
T_{Ant}^{iono}(\nu,TEC(t);\Theta_0,\Phi_0)&=&T_{sky}^{iono}(\nu,t;\Theta_0,\Phi_0)(1-\tau (\nu,TEC(t))) \nonumber \\
& &   + \tau (\nu,TEC(t))*<T_e>
\label{eq:rte}
\end{eqnarray}
where $T_{sky}^{iono}$ is the modified sky brightness temperature due to ionospheric refraction given by equation~\ref{eq:iono_refrac}, $\tau (\nu,TEC)$ is the corresponding optical depth of the ionosphere ($TEC=\int n_e(s) ds$) given by equation~\ref{eq:relation}, $<T_e>$ is the average thermodynamic temperature (or electron temperature) of the ionosphere causing the thermal radiation, $n_e$ is the electron density in the ionosphere and ($\Theta_0,\Phi_0$) are pointing centers (see equation~\ref{eq:iono_refrac}). $T_{Ant}^{iono}$ is the effective brightness temperature of the trans-ionospheric signal recorded by any ground-based antenna. This signal has been affected by all three ionospheric effects: refraction, absorption and emission. It should be noted here that $T_{Ant}^{iono} (\nu,TEC(t);\Theta_0,\Phi_0) = T_{sky}(\nu,\Theta_0,\Phi_0)$ (see equation~\ref{eq:sky}). In the rest of the section, we will discuss these three effects in details.

\begin{figure*}[t!]
\centering
\epsfig{file=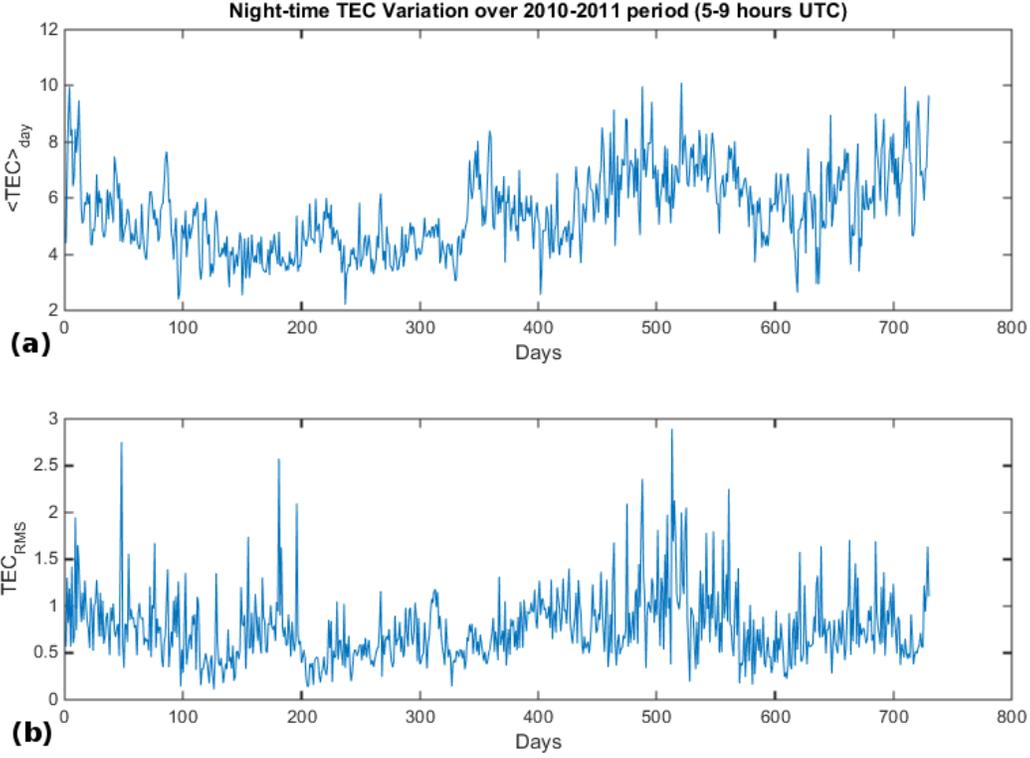,height=4.0truein}
\caption{GPS-derived night-time TEC variation over 2010--2011 period near Green Bank, WV, USA. {\bf (a)} Mean TEC value for each night (for 4 hours night-time data) over different nights for 2010-2011. {\bf (b)} RMS of the night-time TEC values over different nights during 2010 and 2011. This period is near the last Solar Minimum around year 2009. The GPS-TEC data used in these plots have been obtained from the Madrigal database for the ``World-wide GPS Network'' \citep{rideout06}.}
\label{fig:TEC_var_GB}
\end{figure*}
\begin{figure*}[t!]
\centering
\epsfig{file=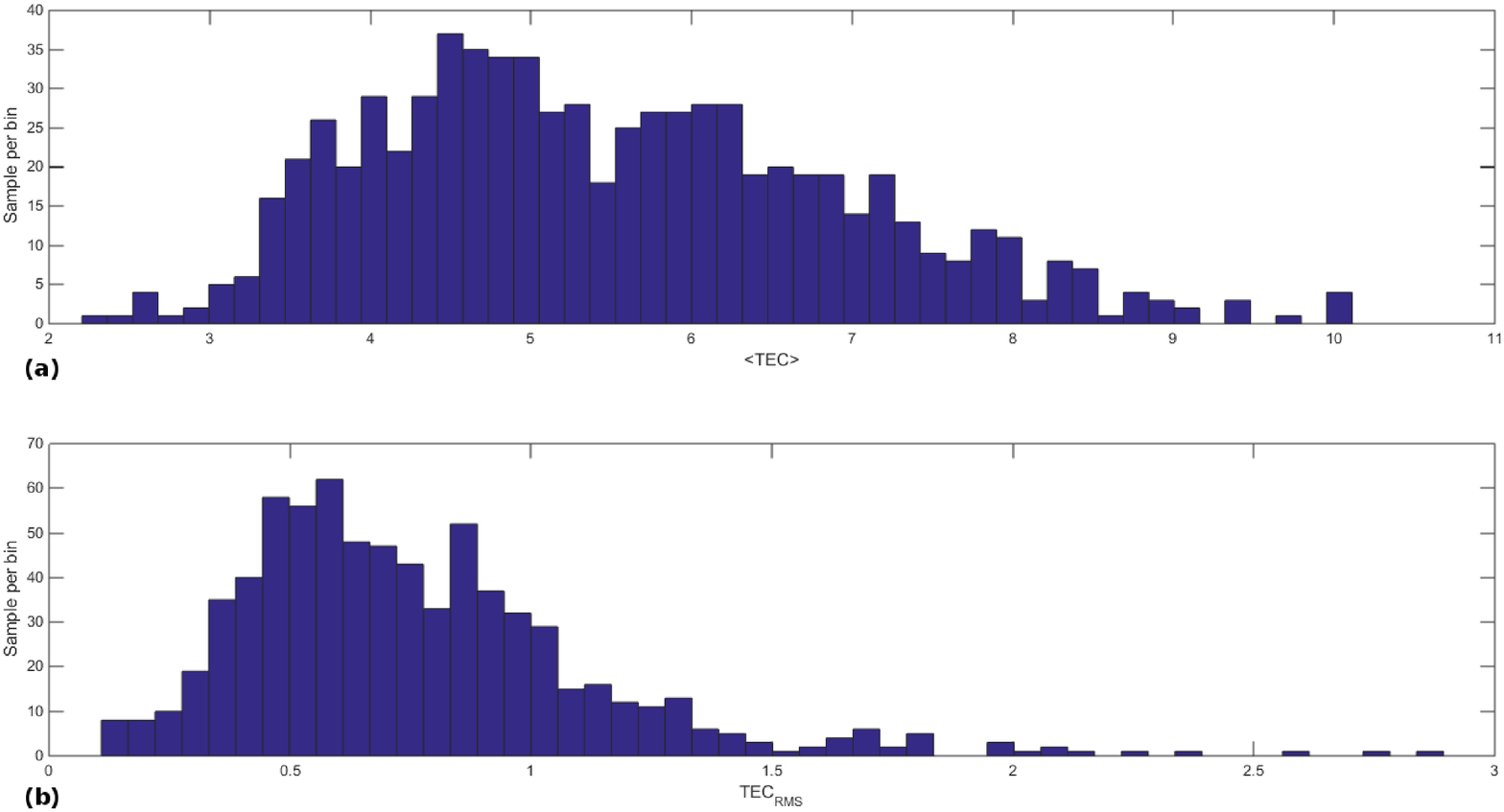,height=2.4truein,width=3.2truein}
\epsfig{file=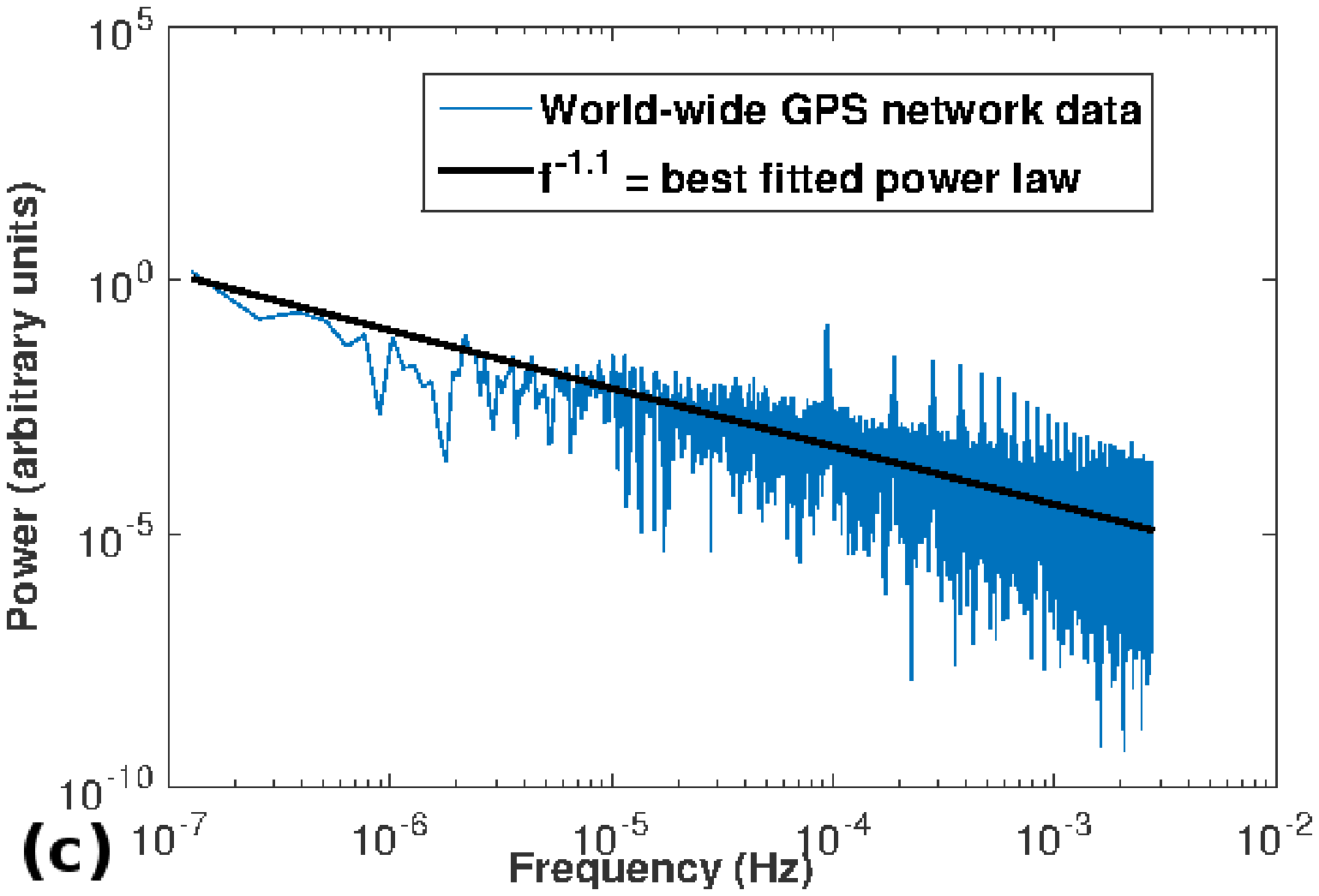,height=2.4truein,width=2.6truein}
\caption{{\bf (a)} Distribution of the mean-TEC values in the night-time over Green Bank, WV, USA for the period 2010-2011. {\bf (b)} Distribution of the RMS of the mean-subtracted TEC values over the same period.  The GPS-TEC data used in these plots have been obtained from the Madrigal database for the ``World-wide GPS Network'' \citep{rideout06}. {\bf (c)} Power spectrum of the night-time TEC variation over GreenBank,WV. The original data are not shown in this paper. However, Figure~\ref{fig:TEC_var_GB} shows the 4 hour night-time mean and RMS of the TEC data over GreenBank. The power is in arbitrary linear units. The x-axis denotes dynamical frequency in Hz (this is the Fourier conjugate of time and should not be confused with the RF frequency of observations). We have also fitted a power law curve to this power spectrum yielding power $\propto 1/f^{1.78}$ (shown in black).}
\label{fig:TEC_var_hist_GB}
\end{figure*}
\begin{figure*}[t!]
\centering
\epsfig{file=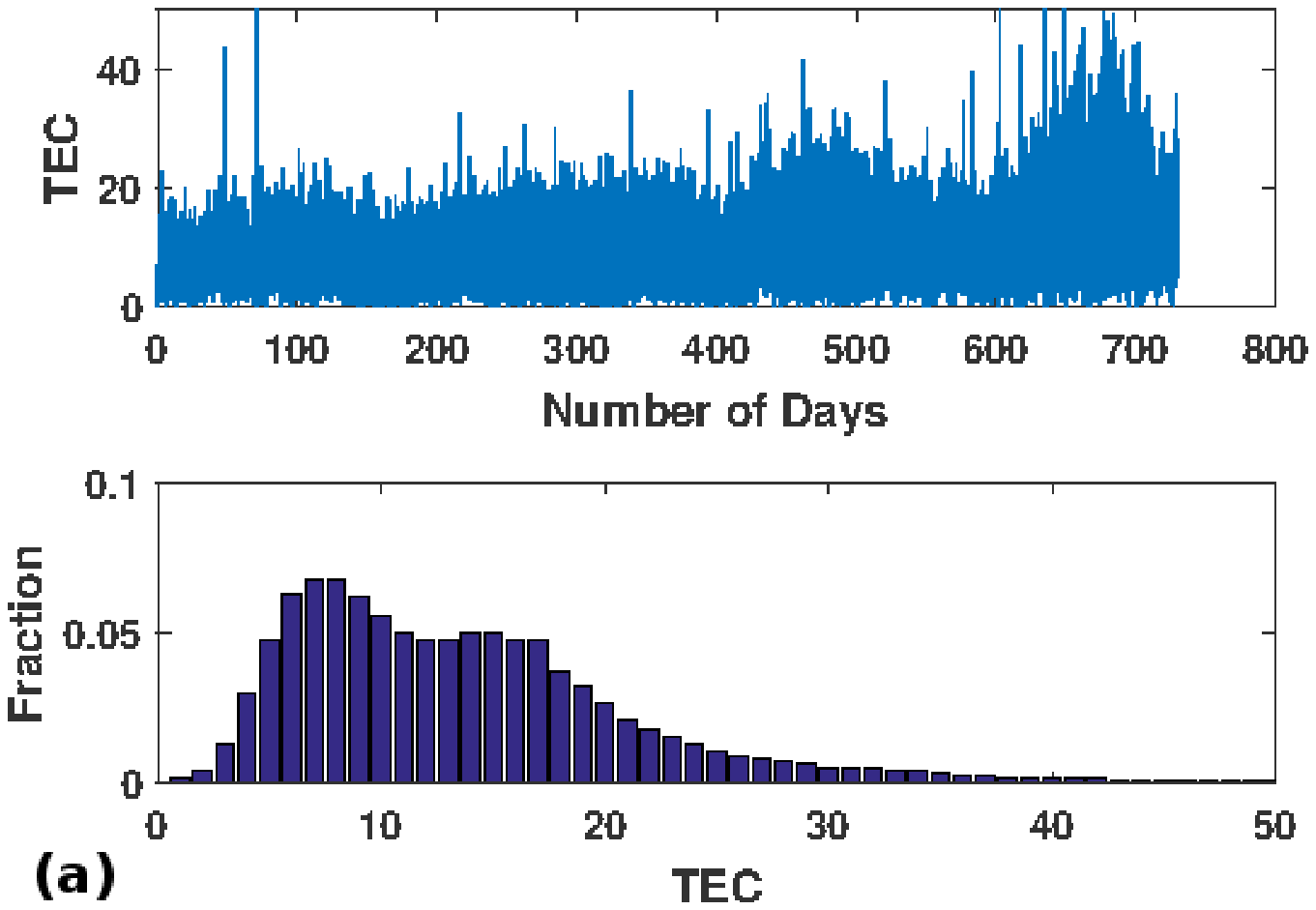,height=2.4truein,width=3.2truein}
\epsfig{file=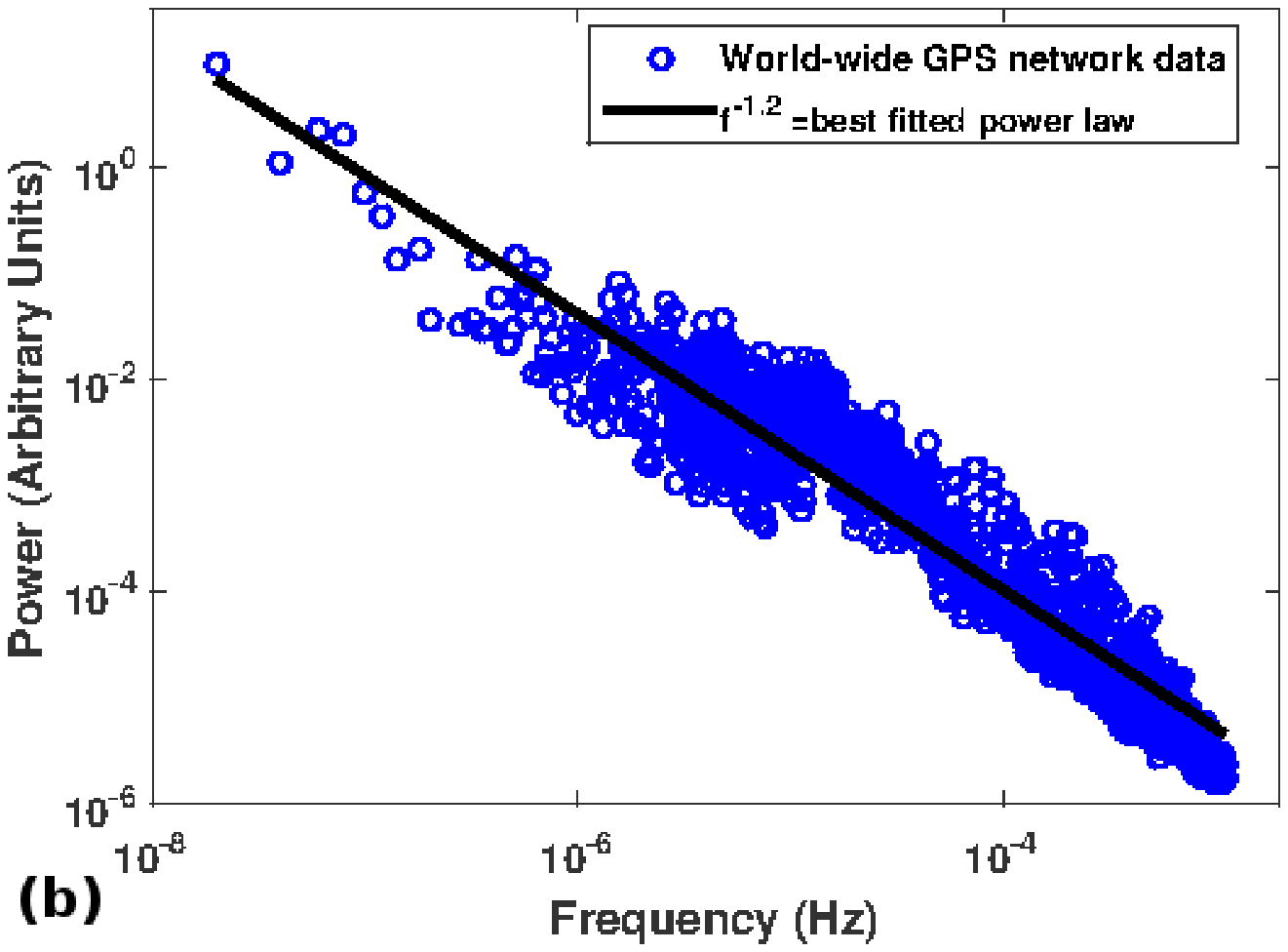,height=2.4truein,width=2.6truein}
\caption{{\bf (a)} Variation in the TEC values in the continuous day+night-time over Green Bank, WV, USA for the period 2010-2011. Distribution of the TEC values over the same period. The GPS-TEC data used in these plots have been obtained from the Madrigal database for the ``World-wide GPS Network'' \citep{rideout06}. {\bf (b)} Power spectrum of the continuous day+night-time TEC variation over GreenBank,WV. The power is in arbitrary linear units. The x-axis denotes dynamical frequency in Hz. We have also fitted a power law curve to this power spectrum yielding power $\propto 1/f^{1.2}$ (shown in black).}
\label{fig:TEC_var_hist_GBday}
\end{figure*}

\begin{figure*}[t!]
\centering
\epsfig{file=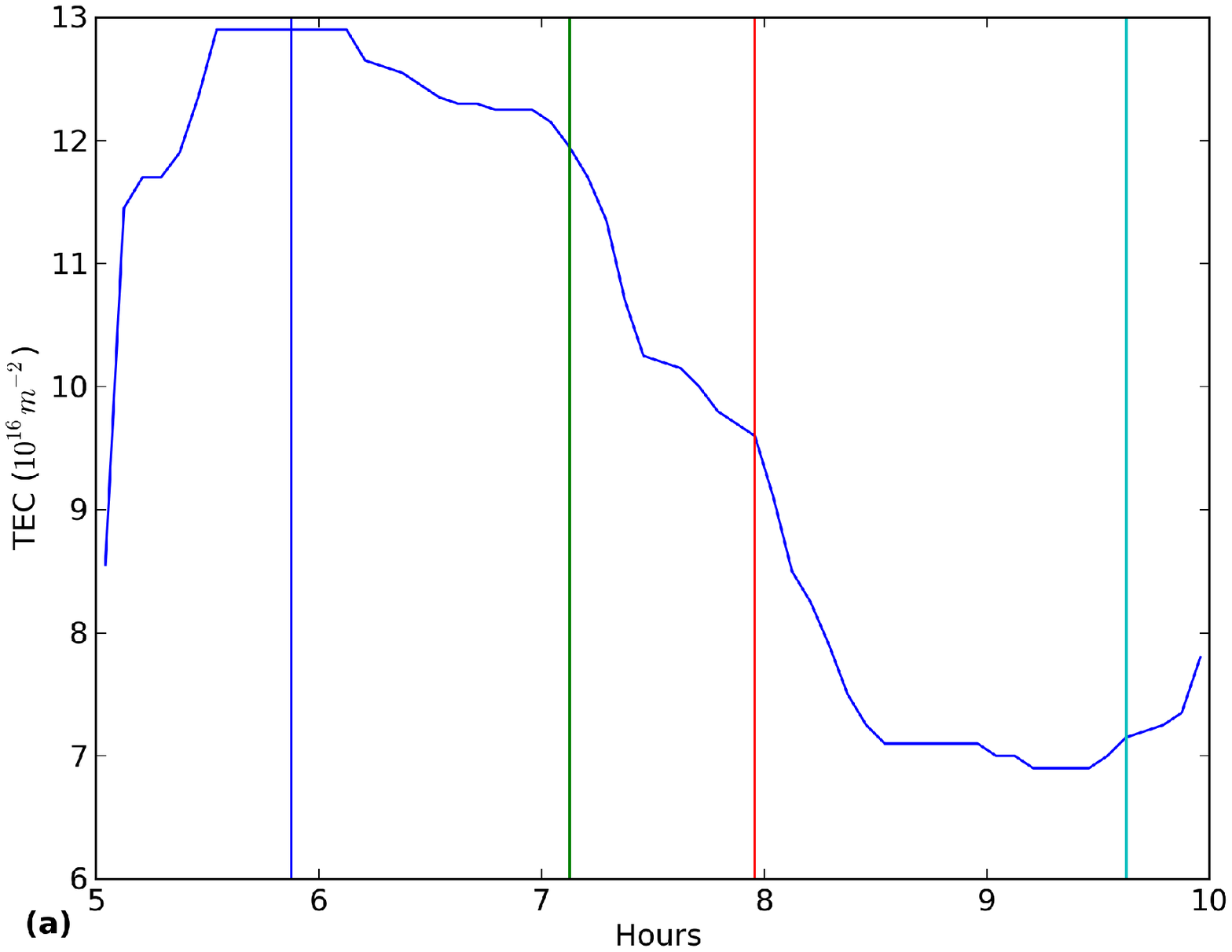,height=2.4truein,width=2.6truein}
\epsfig{file=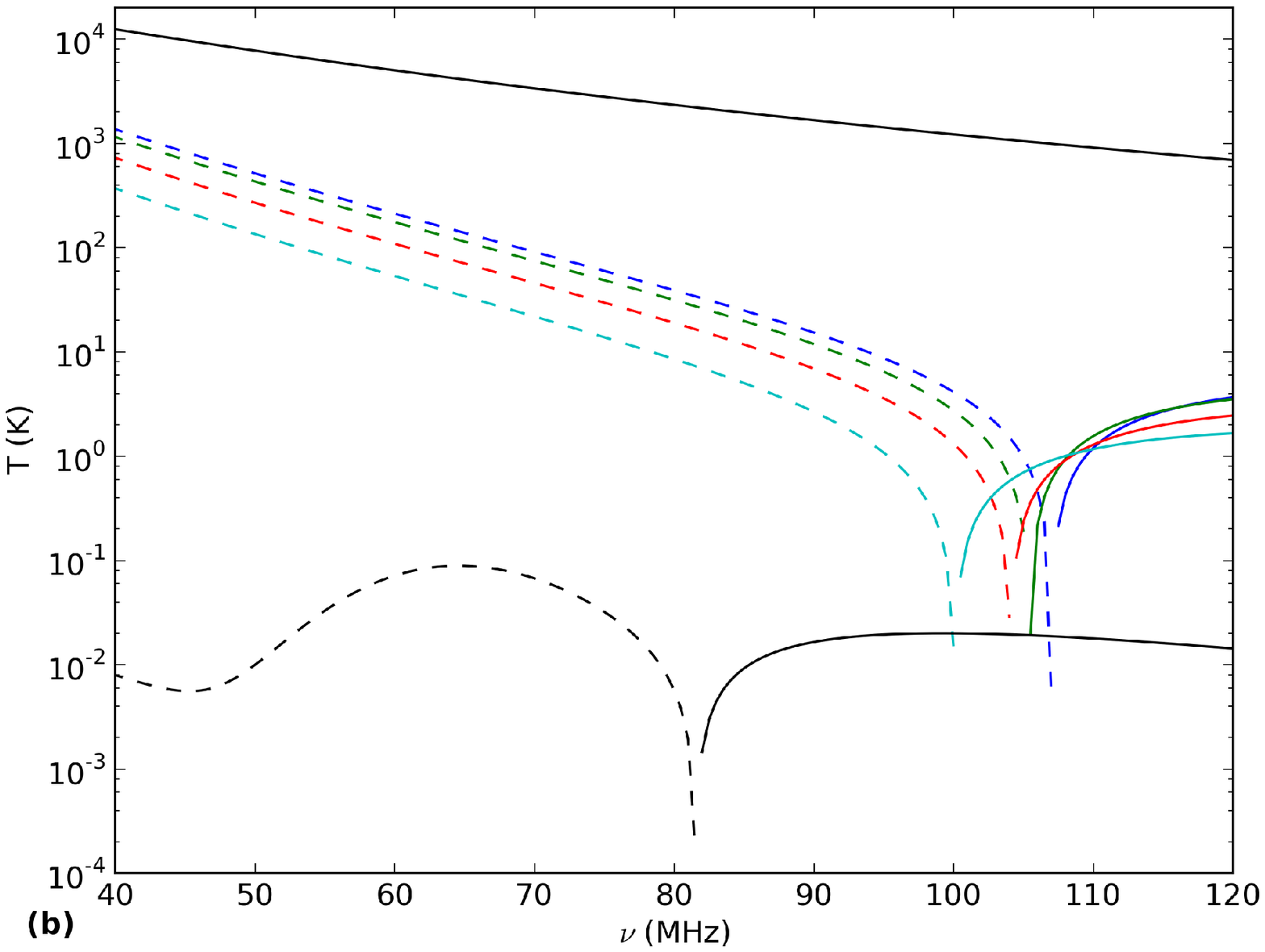,height=2.4truein,width=2.6truein}
\epsfig{file=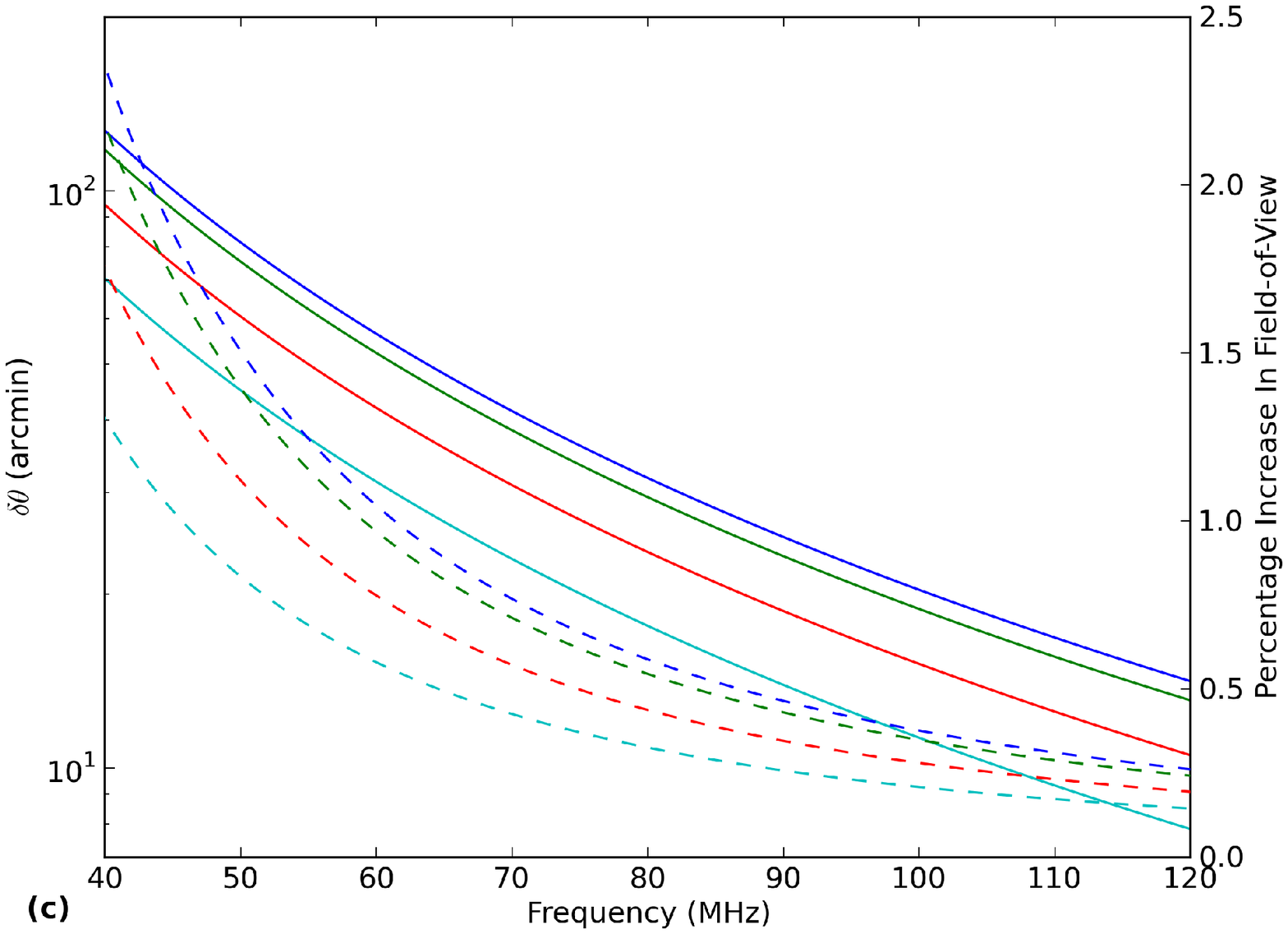,height=2.4truein,width=2.6truein}
\epsfig{file=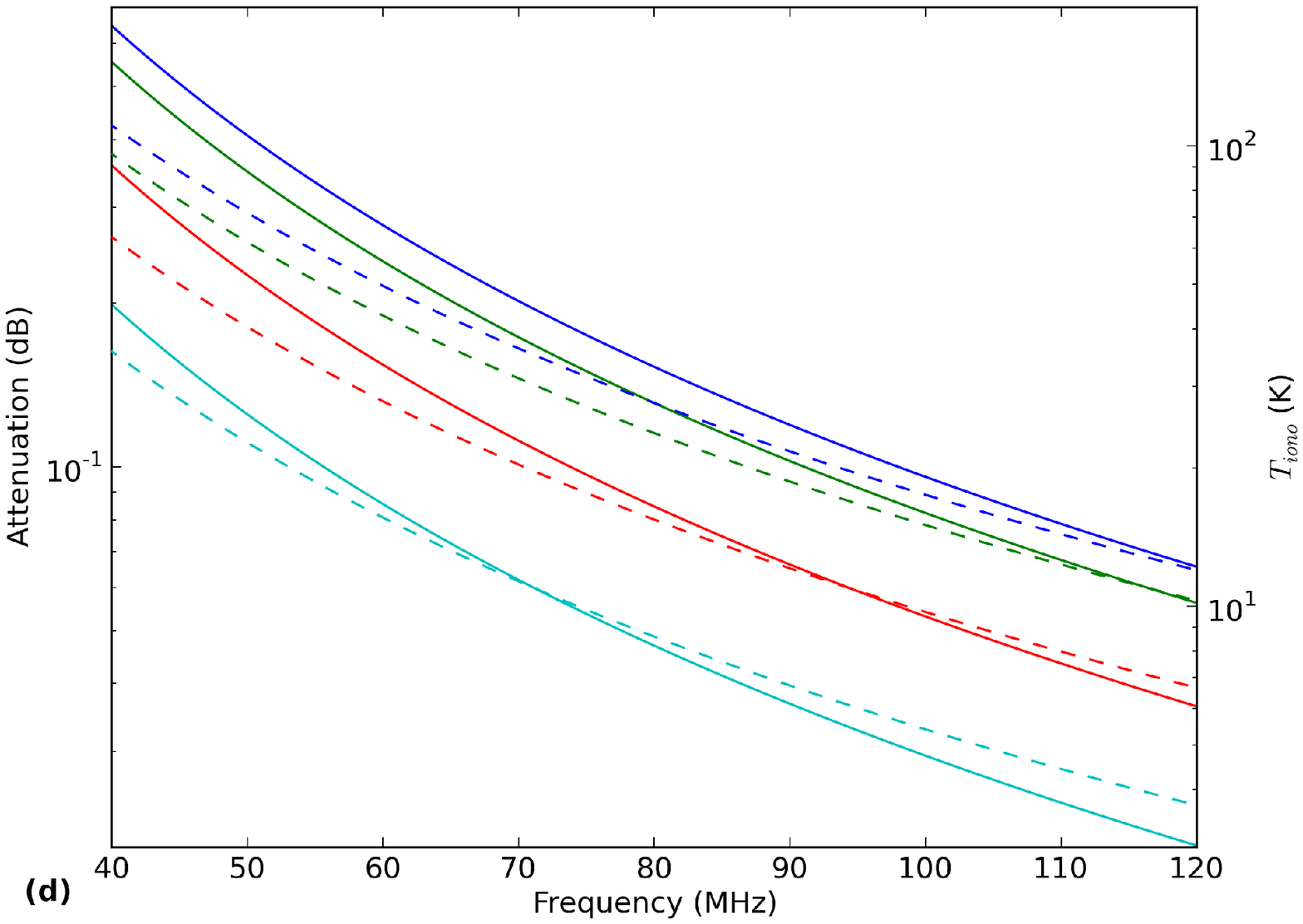,height=2.4truein,width=2.6truein}
\caption{{\bf (a)} GPS-derived TEC variation over Green Bank, WV, USA for a typical night (day 488) between 2010-2011 when the TEC values are relatively high. The vertical 'blue', 'green', 'red' and 'cyan' lines denote 4 time-stamps over this typical night in order to capture the variation in the TEC values. {\bf (b)} The 'blue', 'green', 'red' and 'cyan' lines denotes the residual foreground spectra when the original Global Sky Model (in solid black) is subtracted from the ionosphere-corrupted GSM for the four time-stamps described in the previous plot. Also shown in the global 21 cm signal in black (solid and dashed). The dashed part of the lines denote negative values in respective spectra. {\bf (c)} The deviation angle $\delta \theta$ is plotted (in solid lines) as a function of frequency for the 4 different time-stamps (same colors are used for the respective vertical lines in Figure (a)) over this typical night. Also shown is the variation of the percentage increase in the field-of-view (in dashed lines) over time and frequency. {\bf (d)} Attenuation (in dB) is plotted as a function of frequency (in solid lines) for the 4 different TEC values in Figure (a). Also shown, are the variation in the thermal emission from the ionosphere (in dashed lines).}

\label{fig:TEC_var_hi}
\end{figure*}

\subsection{Refraction}
\label{sec:refrac}
Any incident ray from any part of the sky is refracted as it propagates through the changing density layers of the ionosphere.  Due to its density, the majority of the refraction occurs in the F layer. The refraction at the F-layer of the ionosphere can be compared to a spherical lens where the refracted ray is deviated towards the zenith \citep{harish14}. Due to this refraction, any ground-based radio antenna records signal from a larger region of the sky resulting in excess antenna temperature. 

In order to model the effect of refraction of radio waves in the F-layer, we follow the treatment in \citet{bailey48}. The refractive index ($\eta$) of a radio wave at frequency $\nu$ is given by \citet{bailey48,evans68}:
\begin{equation}
\eta^2(\nu,t) = 1 - \left(\frac{\nu_p (t)}{\nu}\right)^2 \left[1-\left(\frac{h-h_m}{d} \right)^2 \right]
\label{eq:rindex}
\end{equation}
where $h$ is the altitude, $h_m$ is the height in the F-layer where the electron density is maximum, $d$ denotes the change in the altitude with respect to $h_m$ where the electron density goes to zero and $\nu_p$ is the plasma frequency given by \citep{sysbook01}:
\begin{equation}
\nu_p^2(t) = \frac{e^2}{4\pi^2 \epsilon_0 m} n_e(t)
\label{eq:plasma}
\end{equation}
where $e$ is the electronic charge, $m$ is the electron mass, $\epsilon_0$ is the dielectric constant of free space and $n_e$ is the ionospheric electron density.
If we assume that the F-layer is a single with parabolic geometry and bounded by free space with $\eta=1$, then the angular deviation suffered by any incident ray with angle $\theta$ with respect to the horizon (Figure~\ref{fig:iono}(a)) is given by \citep{bailey48}:
\begin{equation}
\delta \theta (\nu,t) = \frac{2d}{3R_E}\left(\frac{\nu_p (t)}{\nu}\right)^2 \left(1+\frac{h_m}{R_E} \right) \left(\sin^2{\theta} + \frac{2h_m}{R_E} \right)^{-3/2} \cos{\theta}
\label{eq:refrac}
\end{equation}
where $R_E = 6378$~km is the radius of the Earth. The above equation shows that the ionospheric refraction scales as $\nu^{-2}$. It is also evident that the maximum deviation occurs for an incident angle of $\theta=0$ or the horizon ray. For a given frequency of observations, the field-of-view will be larger than the primary beam of the antenna (Figure~\ref{fig:sig}(b)) due to this ionospheric refraction.  

The intrinsic sky spectrum ($T_{sky}(\nu)$: see equations~\ref{eq:conv} and \ref{eq:sky}) will be affected by the ionospheric refraction as \citep{harish14}:
\begin{eqnarray}
T_{sky}^{iono}(\nu,t;\Theta_0,\Phi_0) & = & \int_0^{2\pi} d\Phi \int_0^{\pi/2} d\Theta B'(\nu,\Theta-\Theta_0 -\delta\theta(t),\Phi) \nonumber \\
& & T_{sky}(\nu,\Theta-\Theta_0,\Phi-\Phi_0) \sin{\Theta}
\label{eq:iono_refrac}
\end{eqnarray}
where ($\Theta_0, \Phi_0$) is the pointing center. $B'(\nu,\Theta-\Theta_0-\delta\theta,\Phi-\Phi_0)$ denotes the increase in the effective field-of-view due to ionospheric refraction and $T_{sky}(\nu,\Theta,\Phi)$ denotes the model sky map. Following the above equation, we can derive the effective field-of-view and resulting increase in antenna temperature for a given foreground model and ionospheric model. 

In order to estimate the percentage increase in the field-of-view, we have computed the ratio of the deviation of the incident ray at $\theta=0$ and the original field-of-view at that frequency of observations. Since Earth's ionosphere is dynamic (see Section~\ref{sec:iono}) the effective increase in the field-of-view will also change with time. Using this increase as a function of time we have derived the effective HPBW of the Gaussian primary beam as a function of time. We have used this time--dependent Gaussian primary beam to convolve with the global sky map \citep{oliveira08}. The resultant sky spectra as a function of time reflects the effect of ionospheric refraction. 

In our simulations, we assume that the electron density is homogeneous across the entire height of the F-layer, the maximum electron density is contributed at $h_m=300$~km and the thickness of the F-layer is $\sim 200$~km. 

\begin{figure*}[t!]
\centering
\epsfig{file=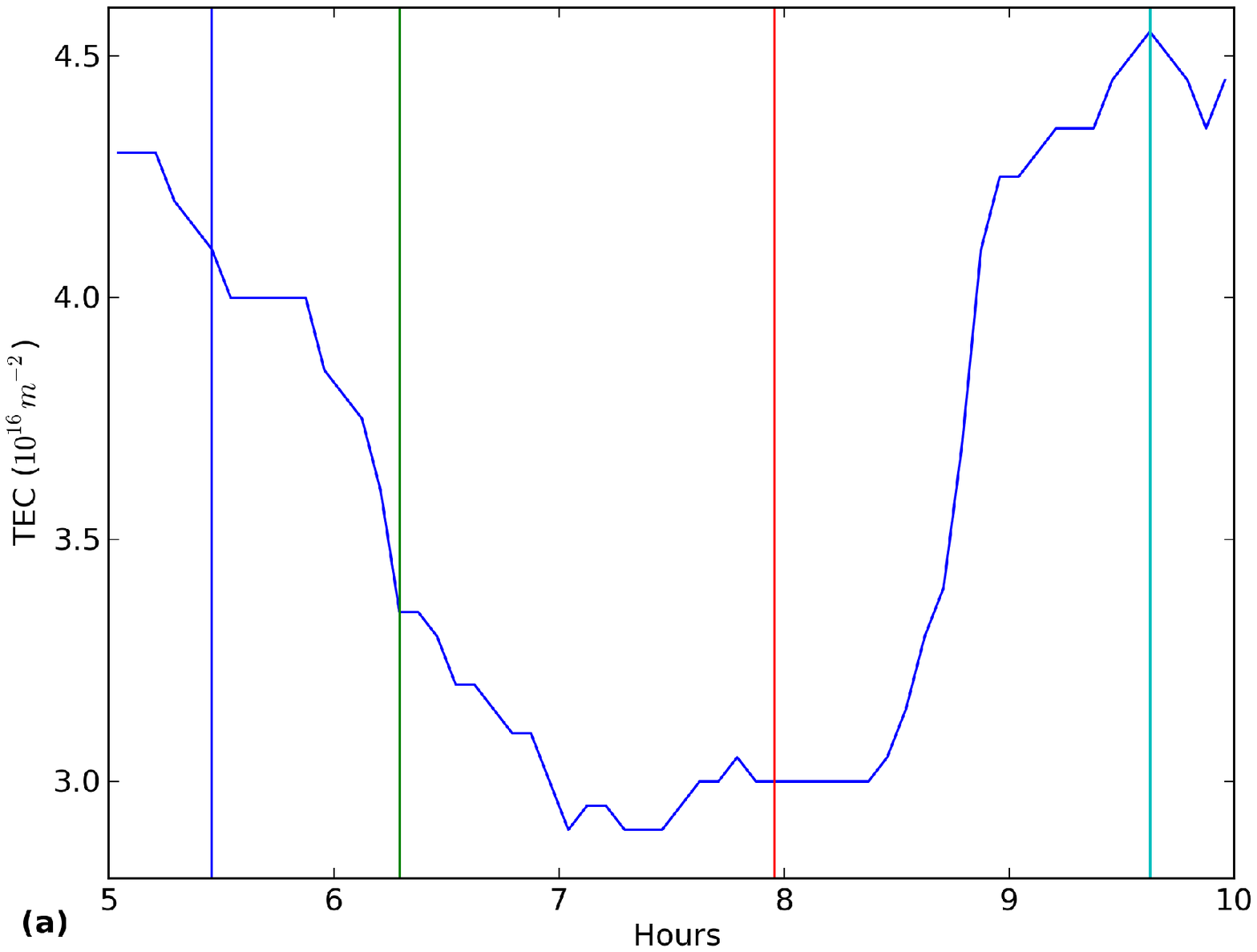,height=2.4truein,width=2.6truein}
\epsfig{file=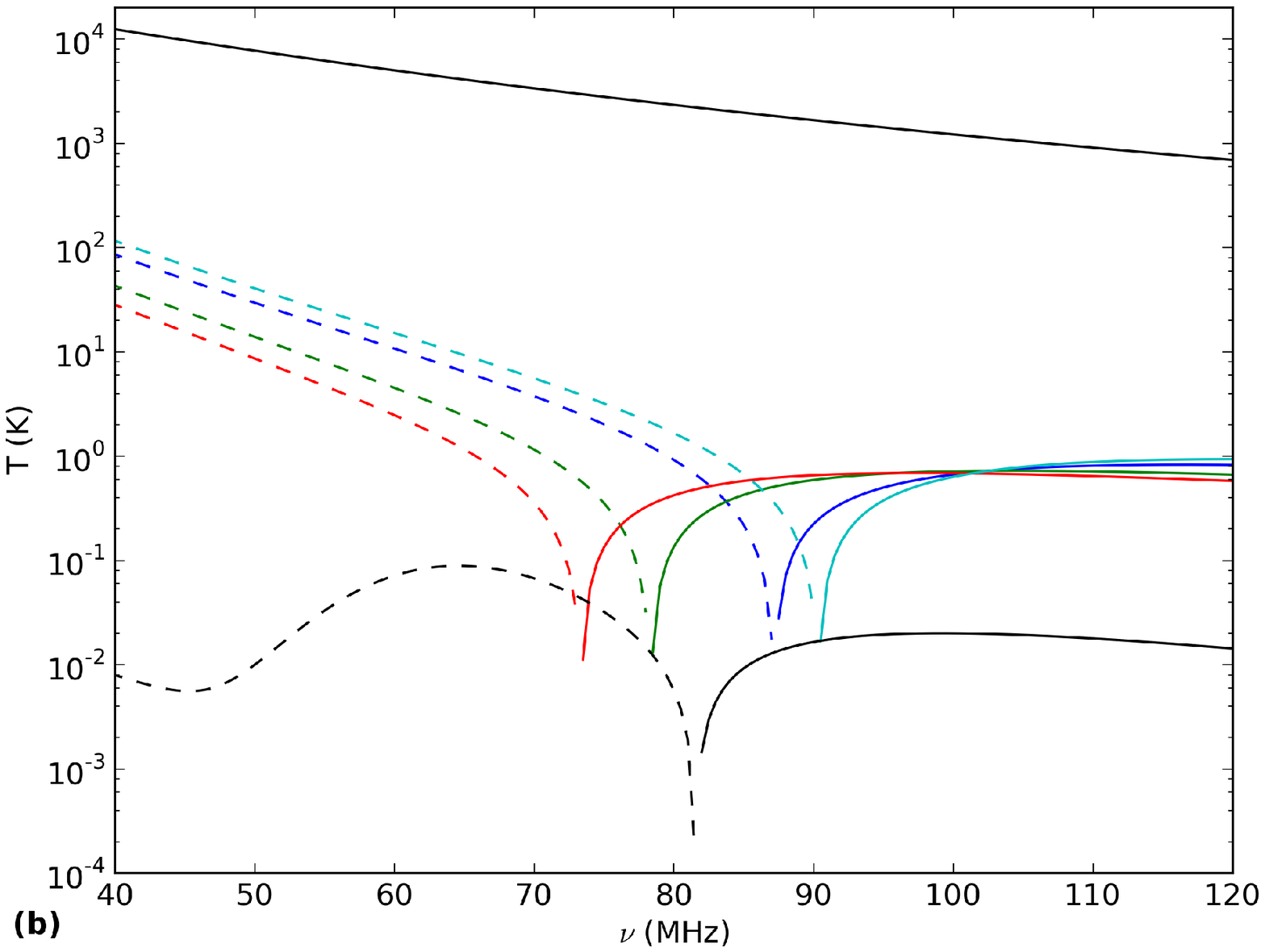,height=2.4truein,width=2.6truein}
\epsfig{file=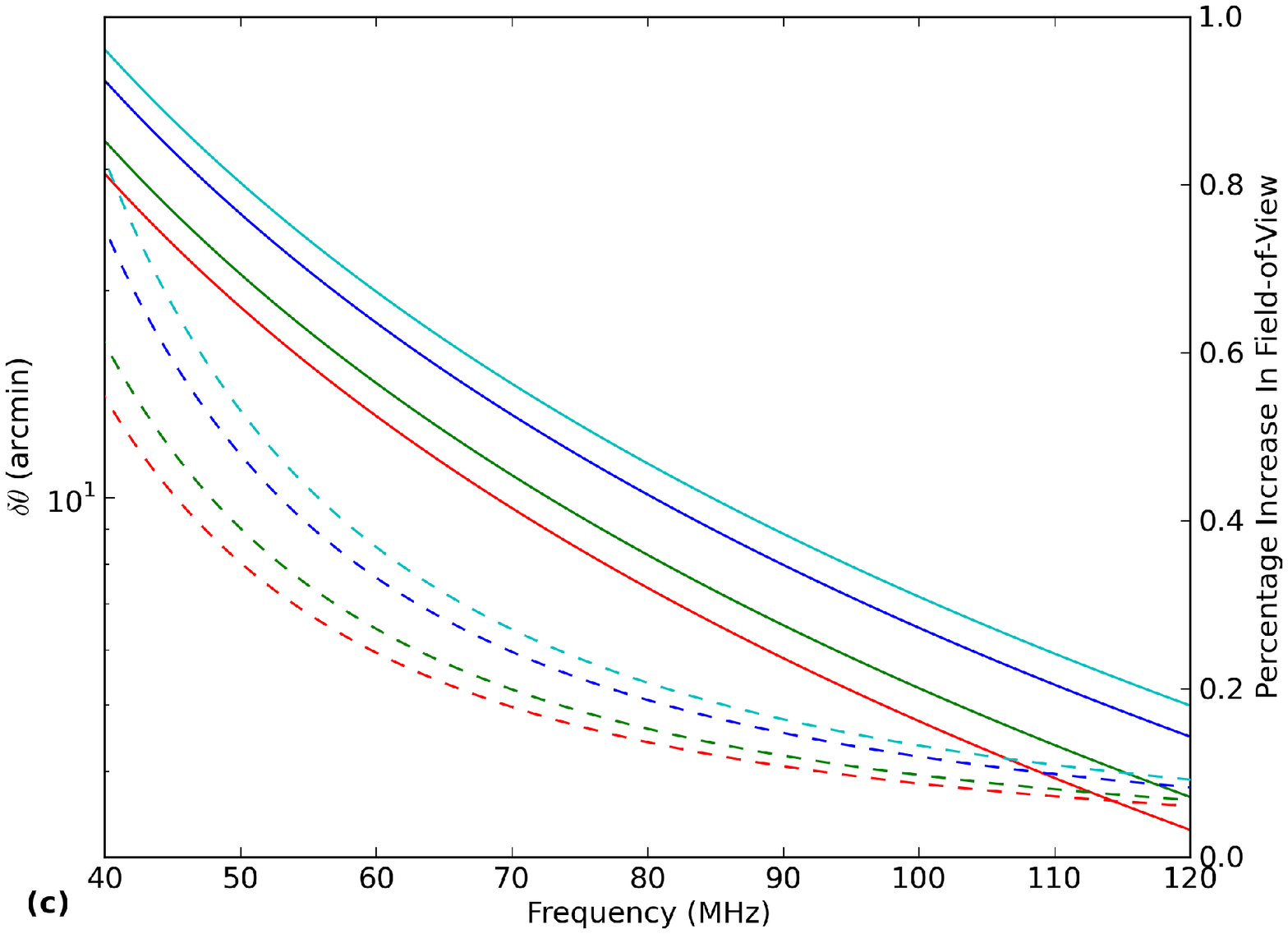,height=2.4truein,width=2.6truein}
\epsfig{file=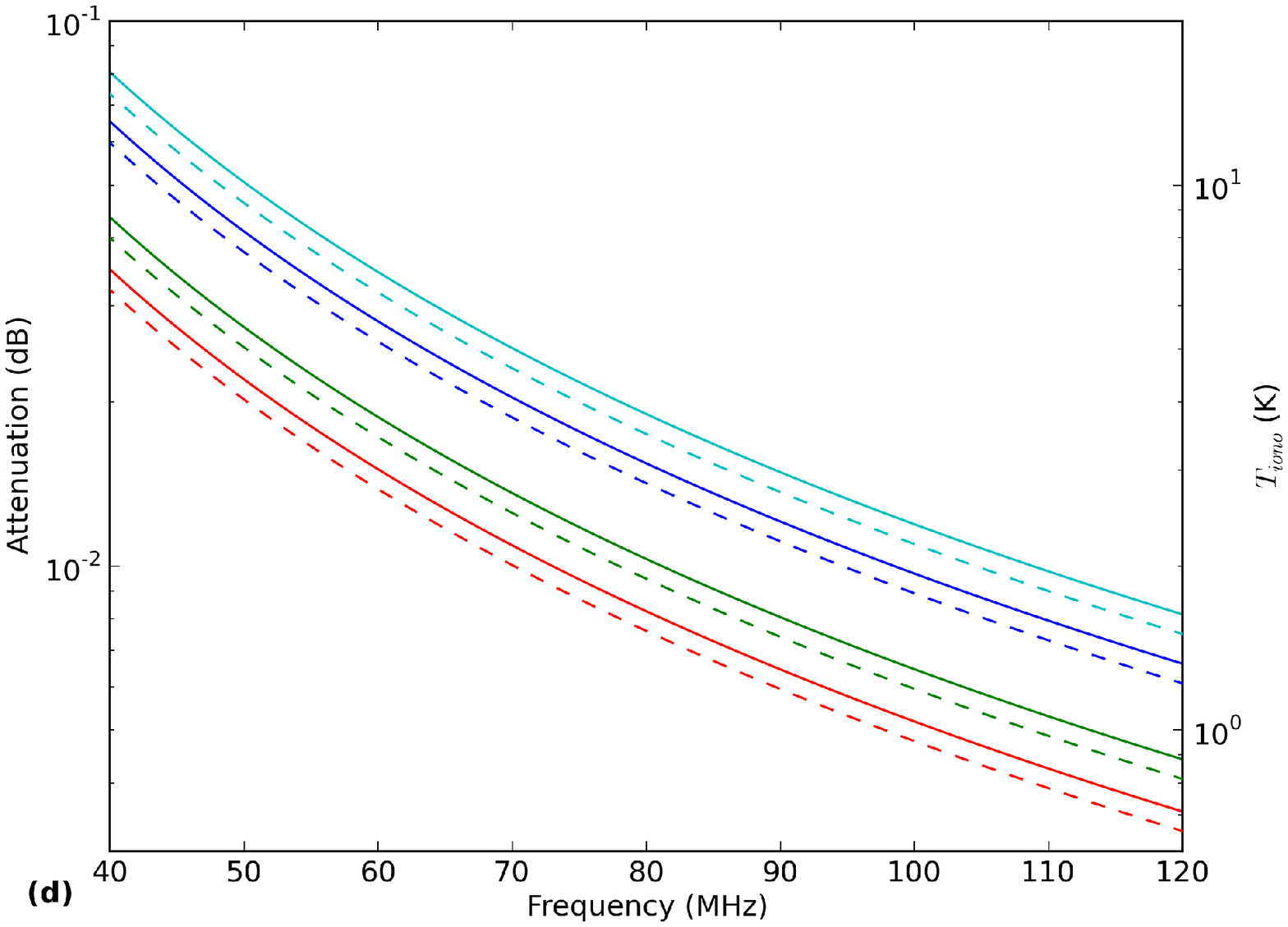,height=2.4truein,width=2.6truein}
\caption{Same as in Figure~\ref{fig:TEC_var_hi} for a typical night (day 198) when the TEC values are relatively low.}
\label{fig:TEC_var_lo}
\end{figure*}

\subsection{Absorption and Thermal Emission}
\label{sec:abs}

The attenuation of the radio waves in the ionosphere is mainly attributed to the D-layer \citep{evans68, davies90}. Total absorption in the D-layer can be expressed in units of dB as \citep{evans68}:
\begin{eqnarray}
L_{dB}(\nu,n_e)  &=& \frac{1.16\times10^{-6}}{\nu^2} \int n_e \nu_c ds ~~ \textnormal{dB} \nonumber \\
L_{dB}(\nu,TEC_D) &=&  \frac{1.16\times10^{-6}}{\nu^2} \langle \nu_c \rangle TEC_D  ~~ \textnormal{dB}
\label{eq:abs}
\end{eqnarray}
where $TEC_D$ is the total electron content (or electron column density) of the D-layer and and $\langle\nu_c\rangle$ is the mean electron collision frequency throughout the ionosphere.  The collision frequency~$\nu_c$ depends upon the local density and is given by \citep{evans68}:
\begin{equation}
\nu_c=3.65 \frac{n_e}{T_e^{3/2}}\left[19.8 + \ln\left(\frac{T_e^{3/2}}{\nu} \right)  \right] ~\textnormal{Hz}
\end{equation}
where $T_e$ is the electron temperature. Generally, the $TEC$ is expressed in units of $1 TECU = 1\times10^{16} \textnormal{m}^{-2}$. From equation~\ref{eq:abs} it is evident that the absorption depends on $\nu^{-2}$. The quantity $L_{\mathrm{dB}}$ is related to the optical depth in equation~\ref{eq:rte} as:

\begin{equation}
L_{dB}(\nu,TEC_D)=10*\log_{10}(1-\tau (\nu,TEC_D))
\label{eq:relation}
\end{equation}
If there is no ionosphere then the $\tau (\nu,TEC_D=0) = 0$ which results in $L_{dB} = 0$.

Apart from absorption, the D-layer is also known to contribute thermal emission \citep{pawsey51,hsieh66a,steiger61} which is given by the final term in equation~\ref{eq:rte}, namely $\tau(\nu,TEC(t)) \langle T_e \rangle$. In our simulations, we have used typical D-layer electron temperature of $T_e=800$~K for mid-latitude ionosphere \citep{zhang04}.

\section{Ionospheric Measurements}
\label{sec:measure}
In the previous section, we have introduced the processes of ionospheric refraction, absorption and emission that affects any trans-ionospheric radio signals. In order to model the effect of Earth's ionosphere on the global 21 cm signal detection from the ground, we need accurate knowledge of: (a) electron densities as a function of height in the D and F layers of the ionosphere and (b) electron temperatures ($T_e$) at the D-layer. The line-of-sight integrated total electron content (TEC) or electron column--density can be derived from the GPS measurements \citep{rideout06,hernandez09,coster12,correia13}, but determination of the electron density as a function of altitude in the ionosphere is highly model--dependent \citep{attila97,bilitza03}. TEC data can be obtained from different GPS measurements for different geo-locations from several GPS-TEC databases (CDDIS IONEX archive\footnote{ftp://cddis.gsfc.nasa.gov/gps/products/ionex/}; \citet{noll10}). In this paper, we have used the GPS-TEC data from the World-wide GPS Network within the Madrigal Database\footnote{http://madrigal.haystack.mit.edu/} \citep{rideout06}. In order to derive the relative contribution of the D-layer and F-layer to the GPS-derived TEC measurements we have used the International Reference Ionospheric model (IRI, \citet{bilitza03}). From the IRI model, we found that the typical ratio between the electron column densities in the D and F layer is about $8\times 10^{-4}$. This value varies by hour of the day, geo-locations and solar activity. Based on the ionospheric conditions over a few chosen sites across the world (see Appendix~\ref{sec:global}), we choose Green Bank, WV as our candidate site to carry out the ionospheric simulations. In this paper, we assume that any ground-based global 21 cm signal observations will only be carried out during the night, when ionospheric effects are smallest.

\begin{figure*}[t!]
\centering
\epsfig{file=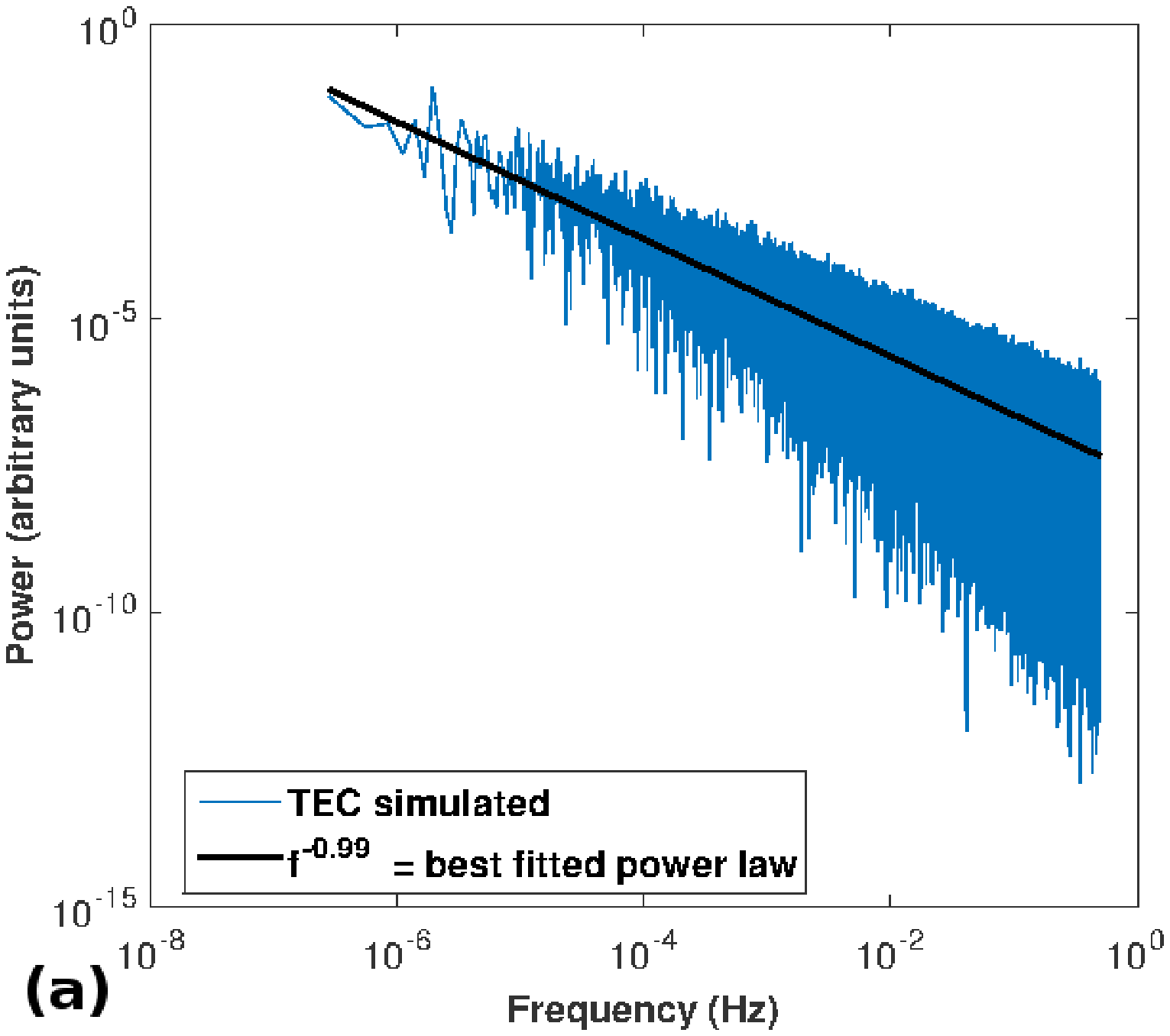,height=2.4truein,width=2.4truein}
\epsfig{file=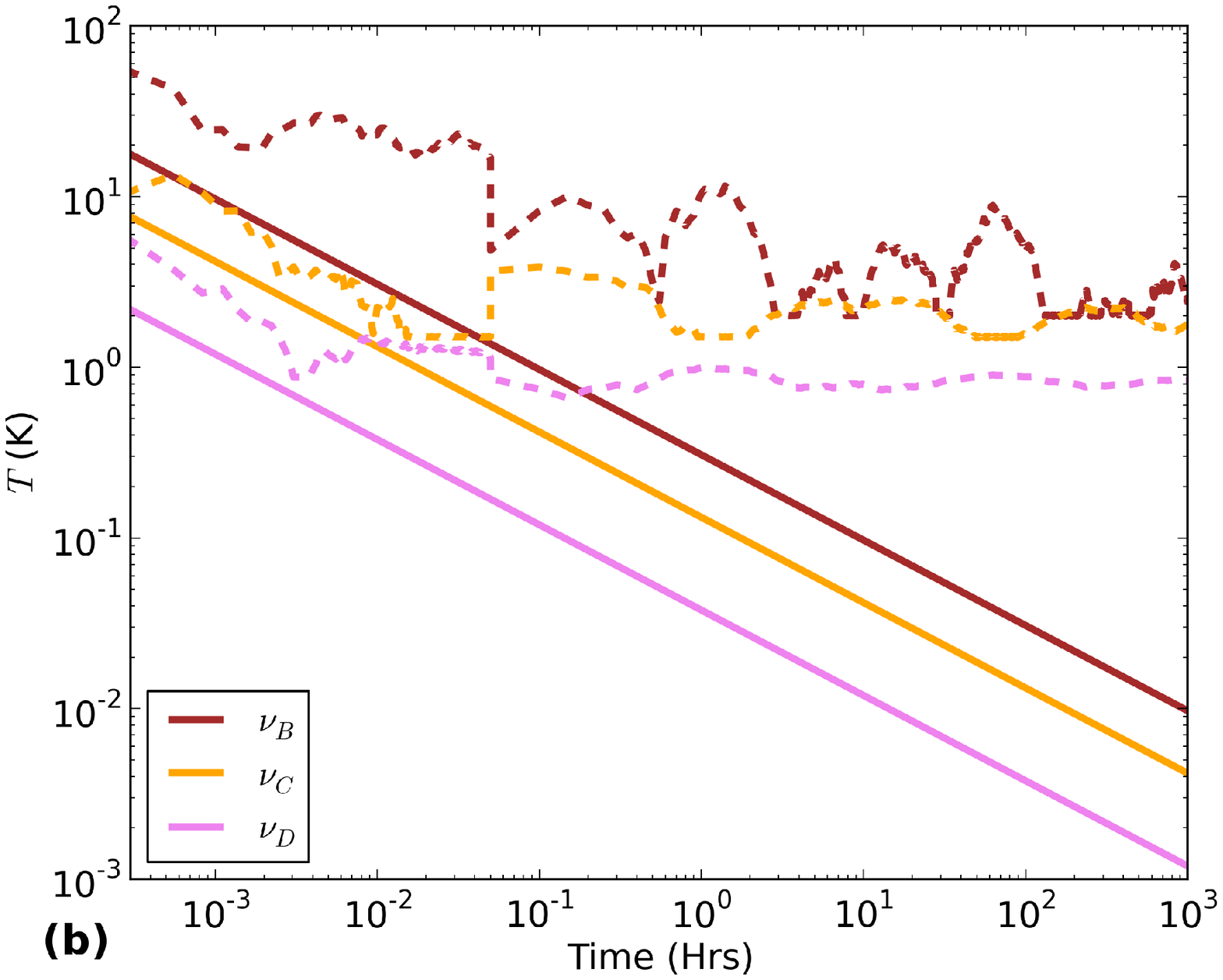,height=2.4truein,width=2.4truein}
\epsfig{file=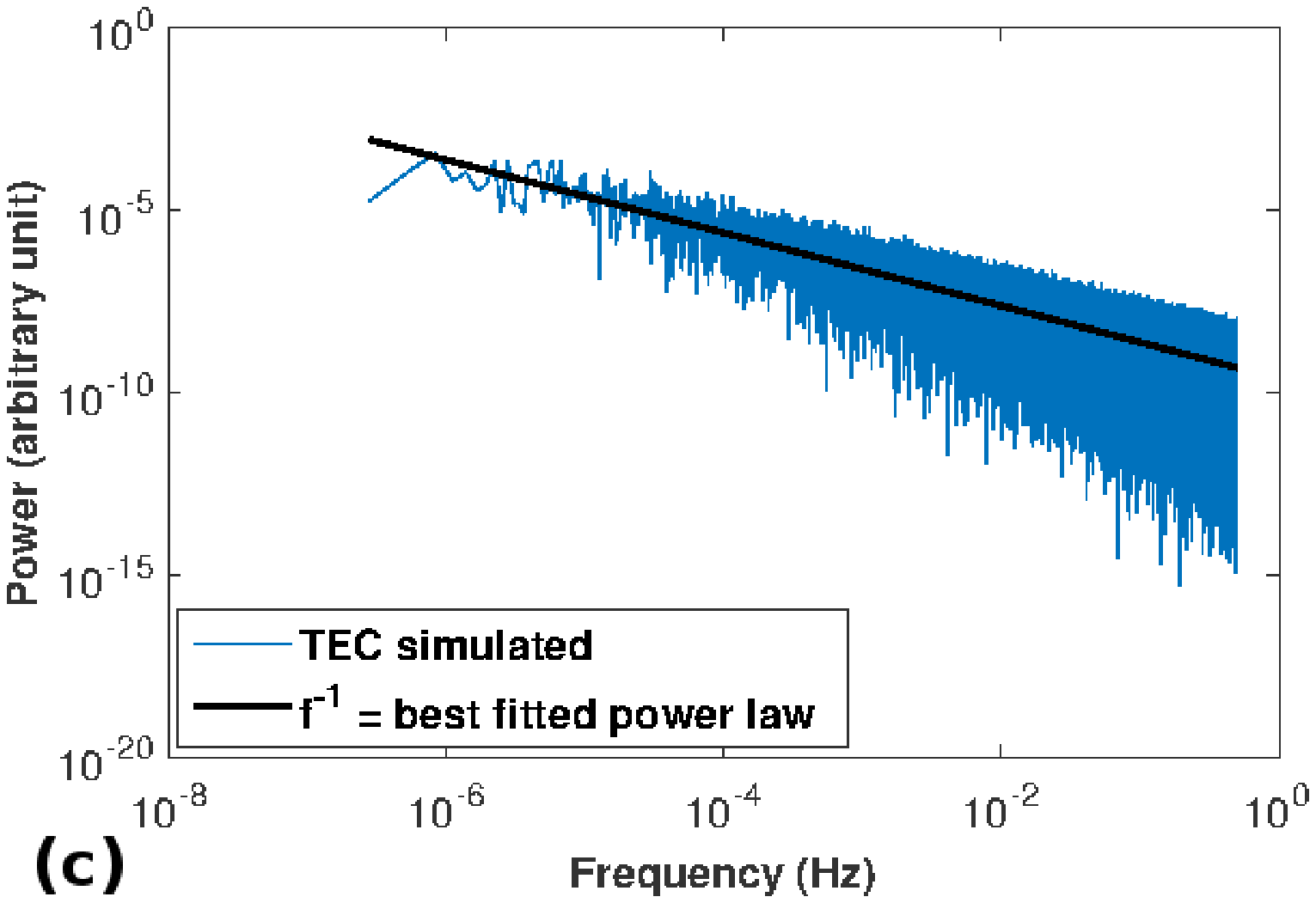,height=2.4truein,width=2.4truein}
\epsfig{file=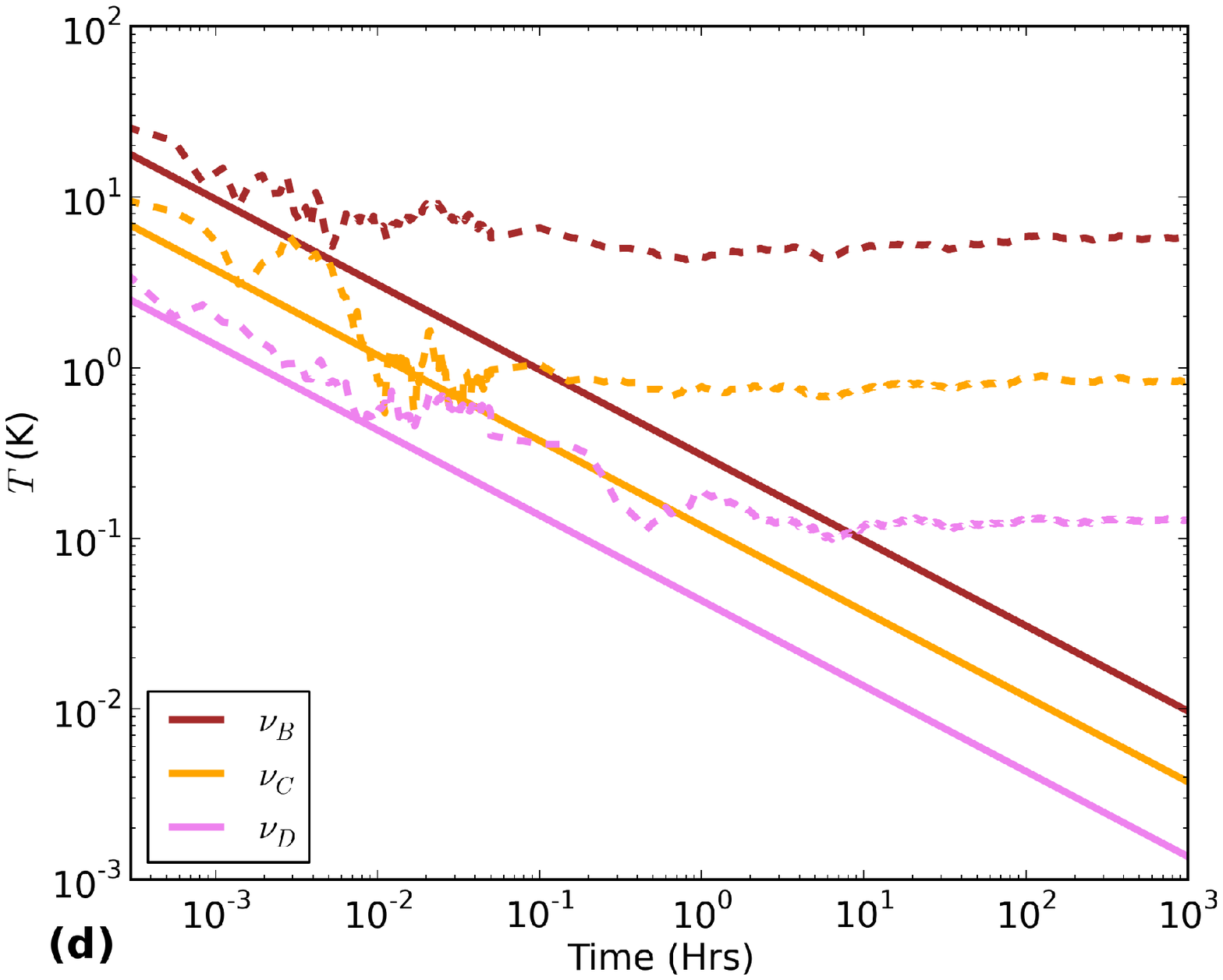,height=2.4truein,width=2.4truein}
\epsfig{file=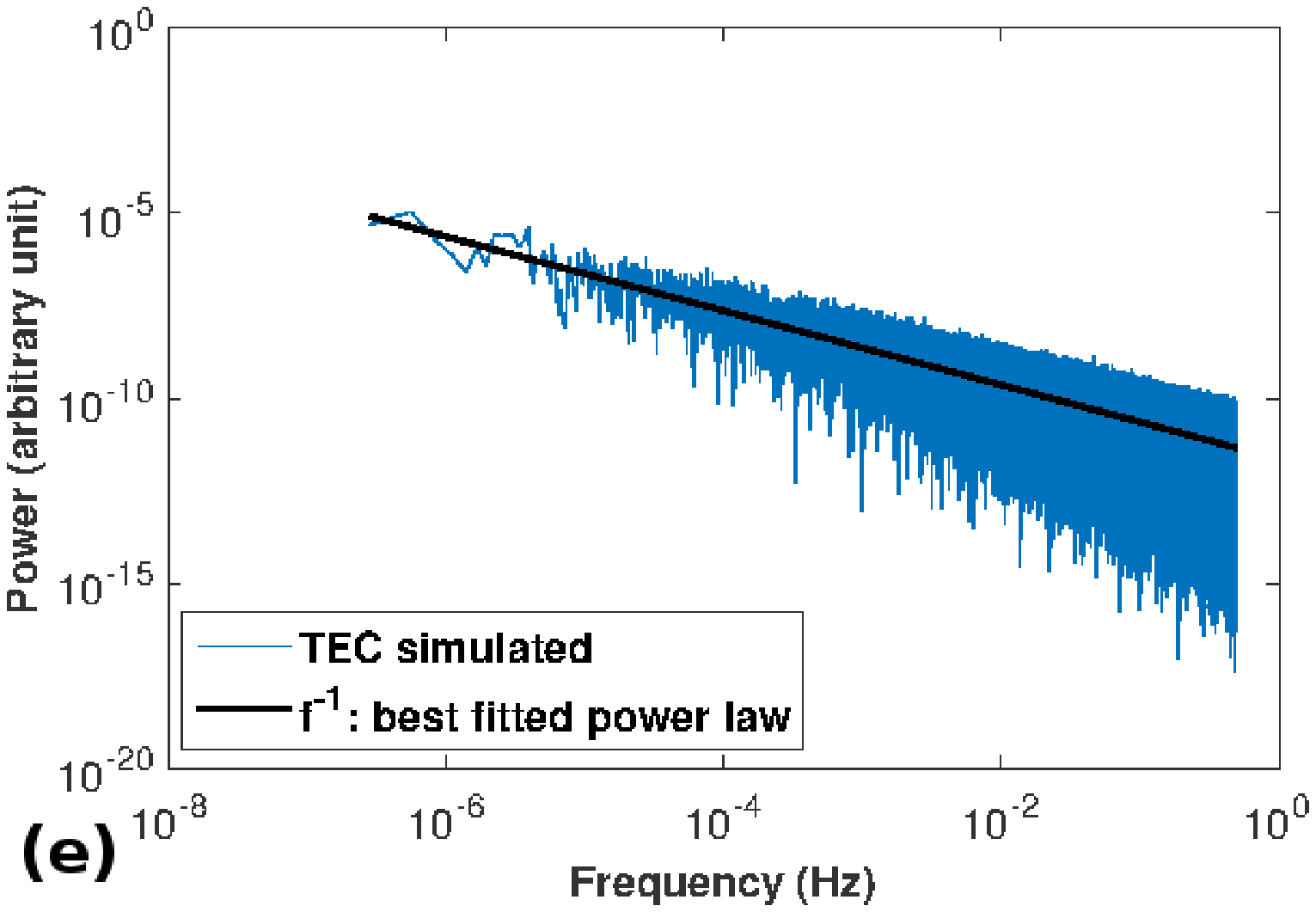,height=2.4truein,width=2.4truein}
\epsfig{file=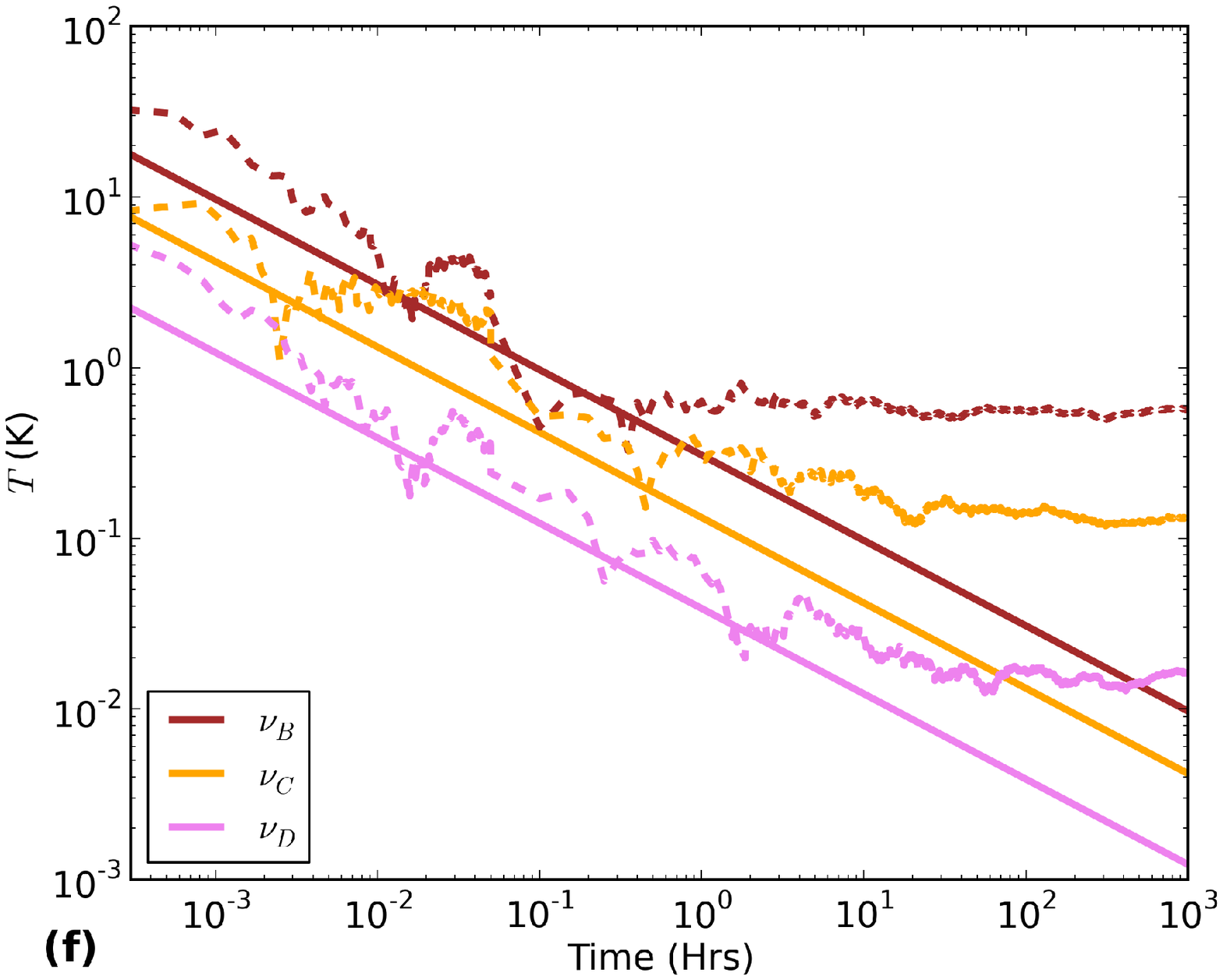,height=2.4truein,width=2.4truein}
\caption{Residual RMS Noise plot for various inaccuracies in ionospheric measurements. {\bf (a)} Power spectrum of the variation in the simulated ionospheric TEC values based on the general night-time TEC values across Green Bank, WV during a solar minimum (Figure:\ref{fig:TEC_var_hist_GB}). The best-fit power law to this power spectrum shows a dependence $\propto 1/f^{1.53}$. This matches with the power spectrum of the actual data taken over Green Bank shown in Figure~\ref{fig:TEC_var_hist_GB}. {\bf (b)} The RMS noise variation (in dashed lines) due to the additional foregrounds created by the ionosphere based on panel (a). The colors brown, orange and magenta denote the location of the turning points B,C and D based on the model 21 cm signal (Figure~\ref{fig:sig}). The solid brown, orange and magenta lines denote the thermal noise variation due to radiometer noise at the same locations of the turning points. The thermal noise added to these simulated data is based on equation~\ref{eq:therm_noise}. {\bf (c)} Power spectrum of the variation in the simulated ionospheric TEC values based on $10\%$ of the normal TEC values across GreenBank, WV. The best-fit power law to this power spectrum shows a dependence $\propto 1/f^{1.62}$. {\bf (d)} Same as in panel (b) but now for ionospheric values from Figure (c). {\bf (e)} Power spectrum of the variation in the ionospheric TEC values based on $1\%$ of the normal TEC values across Green Bank, WV. The best-fit power law to this power spectrum shows a dependence $\propto 1/f^{1.52}$. {\bf (f)} Same as in panel (b) but now for ionospheric values from panel (e).}
\label{fig:iono_rms}
\end{figure*}

\subsection{Effects of Night-time Ionospheric Conditions}
\label{sec:typical}

Figure~\ref{fig:TEC_var_GB}(a) shows variation of the mean night-time (5-9 UTC hours) GPS-TEC values at Green Bank, USA over a 2--year (2010-2011) period near the last solar minimum. The data have a typical time resolution of 15 minutes. Figure~\ref{fig:TEC_var_GB}(b) shows the RMS of the mean-subtracted TEC values ($TEC_{RMS}$) per night over the 2--year period. Figures~\ref{fig:TEC_var_hist_GB}(a) and (b) show the distribution of $\langle TEC \rangle$ and $TEC_{RMS}$. In addition, we have also analyzed the continuous daya and night-time data for these two years. This is presented in Figure~\ref{fig:TEC_var_hist_GBday}. 

It should be noted that such a variation in the ionospheric conditions, where the mean is changing over time along with the variance is again consistent with the ionospheric fluctuations being a flicker noise \citep{wilmshurst90,schmid08}.  In addition, Figure~\ref{fig:TEC_var_hist_GB}(c) shows the power spectrum of the electron density fluctuations with time over Green Bank, WV (Figures~\ref{fig:TEC_var_GB}(a) and (b)). The power spectrum of the electron density fluctuation is $\propto 1/f^{1.78}$ for the night-time data. In addition, we have shown the power spectrum of the electron density fluctuation from the entire 24 hours data over these two years (2010-2011) above GreenBank, WV in Figure~\ref{fig:TEC_var_hist_GBday}b. The best-fit power-law is given by $1/f^{1.2}$. Hence, it is shown that the $1/f$ characteristic is preserved in both all-day data as well as night-time only data for the period 2010-2011. The power-law nature of the electron density fluctuation extends from time-scale of $\sim $ minutes to time-scale of $\sim$ years without a break in the power-law. Figure~\ref{fig:iono}(b) shows the electric field power spectrum as observed by the S33 satellite \citep{temerin89}. The slope of the power spectra is similar. The power spectrum varies as $\propto 1/f^{0.6}$ for values of $10$~Hz~$\lesssim f \lesssim 100$~Hz, and varies as $1/f^{2.6}$ for values of $100$~Hz~$\lesssim f \lesssim 2000$~Hz \citep{temerin89}. On the other hand, \citet{elkins69} shows the power spectrum of ionospheric scintillation varying as $1/f^{2.7}$ at  $10^{-2}$~Hz~$\lesssim f \lesssim 1$~Hz. Hence, it can be noted that the ionospheric activity is composed of different $1/f^\alpha$ processes where $0<\alpha\lesssim 2.5$. The variation in the value of $\alpha$ depends on which layer of the ionosphere is probed during the observations as well as the geo-location and time of the observations with respect to the solar cycle \citep{davies90,roux11}. Comparing the ionospheric observations with the power spectrum of electron density fluctuation as obtained from the GPS data, we can infer that the GPS-TEC data have a $1/f^\alpha$ characteristics where the value of $\alpha$ is within the range of values obtained from other ionospheric measurements \citep{elkins69,temerin89}. Recently, \citet{sokolowski15b} presented their results from BIGHORNS experiment and confirmed that there is a $1/f$ nature in the night-time electron density fluctuation in the ionosphere at frequencies 80-85 MHz. However, their analyses show that the $1/f$ nature suffers a break beyond the timescales of a day. This is in contrary to our findings in Figure~\ref{fig:TEC_var_hist_GBday}. One of the reasons for this discrepancy can be the fact that \citet{sokolowski15b} has used only night-time data which have prevented them to capture the trend beyond few hours in the power spectrum analysis. Hence, the low-frequency break in the power spectrum of the ionospheric electron density fulctuation can be just an artifact of this.

The ``1/f'' noise or flicker noise is a non-stationary random process suitable for modeling time variability of basic parameters of evolutionary systems \citep{keshner82} like solar activity, quasar light curves, electrical noise spectra in devices, ocean current velocity components, fluctuations of the loudness in music, etc. \citep{press78,wilmshurst90,schmid08}. These $1/f^\alpha$ processes create non-Gaussian errors which are independent of the total integration time (see Appendix~\ref{sec:pink}). Hence, the additional noise introduced by the ionospheric effects will not integrate down with longer observations. This non-Gaussian behavior will bound the accuracy at which the composite foreground flux can be measured, and the extent to which it can be effectively removed from the total sky brightness to extract the faint global 21 cm signal.

We now illustrate the effects of ionospheric variations such as those shown in Figure 3 on the extraction of the gloabl 21 cm signal.  We have chosen 2 typical nights: (a) day 488 when the night-time TEC varied between 3 and 16 TECU (Figure~\ref{fig:TEC_var_hi}(a)), and (b) day 198 when the night-time TEC was relatively high, varying between 2.0 and 5.5 TECU (Figure~\ref{fig:TEC_var_lo}(a)). With the values of the GPS-TEC measured over the two typical nights (as mentioned above), we simulated the effects of the ionospheric refraction, absorption and emission in the presence of a foreground sky model (equation~\ref{eq:conv}). 

(1)\textit{Refraction}: Figures~\ref{fig:TEC_var_hi}(c) and \ref{fig:TEC_var_lo}(c) show the change in the deviation angle (for incidence angle $\theta=0$ or horizon ray) and percentage increase in field-of-view due to ionospheric refraction from the F-layer for 4 different time-stamps (corresponding to different TEC values) over two typical nights (mentioned in the beginning of Section~\ref{sec:typical}). The values of these two quantities for TEC~$\simeq 10$~TECU are in good agreement with those derived by \citet{harish14}. It should be noted that the previous work by \citet{harish14} only used a static ionospheric model at 10 TECU to study the refraction effect. 

(2)\textit{Absorption}: Figures~\ref{fig:TEC_var_hi}(d) and \ref{fig:TEC_var_lo}(d) show the change in the absorption term (in dB) over two different nights (mentioned in the beginning of Section~\ref{sec:typical}). The attenuation varies between 0.035 dB (for TEC$\sim 3$~TECU) and 0.65 dB (for TEC$\sim 13$~TECU) at 40 MHz. Typical night-time attenuation varies from 0.05-0.3 dB at 100 MHz \citep{evans68} for the D-layer. Our results are consistent with these observations at 100 MHz. However, the F-layer also contributes to the absorption \citep{shain54, ramanathan59, fredriksen60, steiger61} which currently has not been taken into account in our simulations. Inclusion of the F-layer absorption will increase the total absorption that a radio signal will suffer due to the ionosphere. Moreover, \citet{harish14} have shown that the attenuation also depends on the incidence angle. The attenuation factor can increase by a factor of $\sim 6-7$ due to changing angle of incidence. Recently, \citet{rogers15} detected the effects of the ionosphere in EDGES observations at 150 MHz. Their results have $\Delta \tau_\nu \approx 1\%$ which translates to a $\Delta L_{dB} (=1-\Delta \tau_\nu)=0.04$~dB at 150 MHz. These values are consistent with our results. This agreement validates the modeling and simulation of the dynamic ionosphere that is performed in this paper.

(3)\textit{Emission}: Figures~\ref{fig:TEC_var_hi}(d) and \ref{fig:TEC_var_lo}(d) also show the change in the thermal emission at four different time-stamps over two typical nights (mentioned in the beginning of Section~\ref{sec:typical}). Thermal emission varies from $\sim 6$~K (for TEC$\sim 3$~TECU) to $\sim 100$~K (for TEC~$\sim 13$~TECU) at 40 MHz. Hence, the thermal emission is not the dominant effect of the ionosphere. However, it should be noted that the variation in the electron temperature $T_e$ cannot be determined from the GPS-TEC measurements and has to be gathered from IRI-like models or from back-scatter radar experiments. So any variation in the electron temperature can potentially affect the detection of the faint global 21 cm signal. Recently, \citet{rogers15} derived the electron temperature from 150 MHz observations with EDGES. Their results show a typical electron temperature of 800 K. All our analysis is based on a fixed electron temperature of 800 K (see Section~\ref{sec:abs}) which is also the typical electron temperature above Green Bank, WV.

(4)\textit{Combined Effect}: Figures~\ref{fig:TEC_var_hi}(b) and \ref{fig:TEC_var_lo}(b) show the combined effect of ionospheric refraction, absorption and emission. The simulated spectra with the combined effect of the ionosphere is given by $T_{obs}(\nu,TEC(t);\Theta_0,\Phi_0)=T_{Ant}^{iono}(\nu,TEC(t);\Theta_0,\Phi_0) + T_n$ where $T_{Ant}^{iono}$ is given by equation~\ref{eq:rte} and $T_n=100$~K is the receiver noise temperature. In addition, the simulated spectra contains the thermal noise given by equation~\ref{eq:therm_noise} where $T_{sys}=T_{Ant}^{iono} + T_n$. The residuals $T_{obs}(\nu,TEC(t);\Theta_0,\Phi_0)-T_{sky}(\nu,\Theta_0,\Phi_0)$ (see equations~\ref{eq:sky} and ~\ref{eq:iono_refrac}) are essentially the additional foregrounds created due to the ionospheric effects. Here, we are demonstrating the effect if we ignore any ionospheric calibration for global signal experiments. Four different TEC values are chosen for each night and are shown in vertical blue,green,red and cyan lines in the Figures ~\ref{fig:TEC_var_hi}(a) and \ref{fig:TEC_var_lo}(a). Corresponding residual spectra are shown in four curves (blue,green,red,cyan) in Figures ~\ref{fig:TEC_var_hi}(b) and \ref{fig:TEC_var_lo}(b). It is evident that the magnitude of these residuals depends on the TEC value for that particular time-stamp as well as on the frequency of observations. The most striking characteristics in these residuals are the ``spectral dips'' in the absolute value of the residuals, which also vary with TEC (or time). These spectral features in the residuals are qualitatively similar to those in the absolute value of the model global 21 cm signal (black, dashed-solid line in Figures~\ref{fig:TEC_var_hi}(b) and \ref{fig:TEC_var_lo}(b)). Such variable spectral features when averaged over long integration time (in actual experiment) will offset the global 21 cm signal from Cosmic Dawn and Dark Ages. Such a non-smooth, time-variable ionospheric foreground will inevitably complicate the extraction of the weak 21 cm signal using the Bayesian routines like Markov Chain Monte Carlo \citep{harker12}, as well as any other approach that works with spectra integrated over long observations affected by the dynamic ionosphere. Hence, even in a typical night with quiet ionospheric conditions (like in Figure~\ref{fig:TEC_var_lo}), the ionospheric effects are major obstacles in the detection of the faint global 21 cm signal.

\subsection{Uncertainties In The Ionospheric Measurements}
\label{sec:errors}
In order to detect the global 21 cm signal, any experiment has to observe for long hours over quiet night-time conditions. The thermal noise in any measurement (see equation~\ref{eq:therm_noise}) reduces ($\propto 1/\sqrt{\delta t}$~ or $1/\sqrt{N_{\textnormal{samples}}}$) for an integration time $\delta t$. However, the additional foreground introduced by ionospheric effects is not noise-like and will not reduce with longer observing time. In Figure~\ref{fig:TEC_var_hist_GB}(a), the mean TEC values over the night-time period in Green Bank varies between $\sim 3-9$ TECU and distribution of the mean-subtracted RMS TEC peaks at $\sim 0.2$ and $1.5$ TECU. This variation in the TEC values reflects the ionospheric variability in the absence of any major solar activity. 
In order to model the effects of the night-time ionospheric variations on total-power observations of the global 21 cm signal, we have considered a mock observation over 1000 hours which is necessary to detect turning point `B' in Figure~\ref{fig:sig}(b) \citep{burns12}. The details of the simulations can be outlined as:
\begin{itemize}
\item Here, we have assumed that care will be taken to remove nights and individual time-stamps with high TEC values and only time-stamps with low TEC values will be retained to extract the global 21 cm signal.

\item We have also assumed that the variation in the low ionospheric TEC values can be represented by a ``1/f'' process where the TEC values represent the usual night-time TEC values above GreenBank, WV during solar minima (Figure~\ref{fig:iono_rms}(a)). We should also note that the power spectra of these synthetic data on TEC variability ($\propto 1/f^{1.53}$) resembles closely the power spectra of the night-time variability of the actual GPS-TEC data ($\propto 1/f^{1.78}$) as shown in Figure~\ref{fig:TEC_var_hist_GB}(c). It should be noted that these values are still lower than the typical variation at Green Bank and mostly reflect the best possible ionospheric conditions that can occur irrespective of the location on the Earth.

\item The simulated spectra with the combined effect of the ionosphere is given by $T_{obs}(\nu,TEC(t);\Theta_0,\Phi_0)=T_{Ant}^{iono}(\nu,TEC(t);\Theta_0,\Phi_0) + T_n$ where $T_{Ant}^{iono}$ is given by equation~\ref{eq:rte} and $T_n=100$~K is the receiver noise temperature.

\item In our simulations, the ionospheric TEC value is chosen from a $1/f$ distribution (mentioned above) every 1 second. The underlying process to create a $1/f$ distribution involves generating a vector of (uniform) random numbers in time series, Fourier transform it, multiply by a weighting factor, and inverse Fourier Transform it back to time domain. The resultant synthetic spectrum $T_{obs}(\nu,TEC(t);\Theta_0,\Phi_0)$ is generated for every time-stamp (i.e. 1 second).

\item In addition, the simulated spectra contains the thermal noise given by equation~\ref{eq:therm_noise} where $T_{sys}=T_{Ant}^{iono} + T_n$.

\item It should be noted that $T_{sky}(\nu,\Theta_0,\Phi_0)=T^{iono}_{Ant}(\nu,TEC=0;\Theta_0,\Phi_0)$ (see equations~\ref{eq:sky} and ~\ref{eq:rte}).

\item Hence, the residuals $T_{obs}(\nu,TEC(t);\Theta_0,\Phi_0)-T^{iono}_{Ant}(\nu,TEC=0,\Theta_0,\Phi_0)$ are essentially the additional foregrounds created due to the ionospheric effects. RMS value of the residuals are calculated over 0.5 MHz channel-widths and plotted in Figure~\ref{fig:iono_rms}(b). 
\end{itemize}

Figure~\ref{fig:iono_rms}(b), shows the RMS value near the locations of the turning points `B' (in blue), `C' (in green) and `D' (in red). The RMS values (in dashed lines) reflect the effect of the additional foregrounds due to the ionosphere. Figure~\ref{fig:iono_rms}(b) also shows the expected reduction in the ideal radiometer noise (equation~\ref{eq:therm_noise}) component with increase in effective observing time. It is evident that even in these low ionospheric conditions, the additional ionospheric foreground does not allow the RMS noise to decrease with time.

From the results in Figure~\ref{fig:iono_rms}(b) it is evident that the effect of the ionosphere on global 21 cm experiments cannot average down with longer observations. Hence, it is critical to calibrate the ionospheric corruption from the global 21 cm data. The accuracy of any such ionospheric calibration will depend on the accuracy of the time-dependent ionospheric parameters like TEC and $T_e$. Currently, the typical errors in the GPS measurements are of the order of $\gtrsim 0.5$ TECU \citep{attila97, hernandez09}. These errors occur due to model-based reconstruction of the vertical TEC from the actual slant TEC measurements as well as other assumptions about the typical ionospheric parameters \citep{attila97}. 

In this paper, we use simulations to understand whether the current or future accuracy of the GPS-TEC measurements will be sufficient to calibrate the ionospheric effects in global 21 cm data-sets, and allow us to detect the spectral features of the global 21 cm signal from the ground. Since the success of any ionospheric calibration depends on the accuracy of the knowledge of the exact ionospheric parameters, we have performed a simulation over 1000 hours' total integration. The procedure of the simulation is mostly similar to that in Figure~\ref{fig:iono_rms}(b). The only changes in this case are:
\begin{itemize}
\item In this case, we have assumed that the simulated spectra is affected by the value of $TEC_{observed}(t) = TEC_{model}(t) + \Delta TEC(t)$, where $\Delta TEC(t)$ denotes the inaccuracy in the ionospheric measurements obtained from GPS.
  
\item $TEC_{model}(t)$ is given by Figure~\ref{fig:iono_rms}(a). $\Delta TEC(t)$ has been randomly chosen every 1 second from a 1/f process shown in Figure~\ref{fig:iono_rms}(c), where the TEC variability is about $10\%$ of that in Figure~\ref{fig:iono_rms}(a). The power spectrum of $\Delta TEC (t)$ (in Figure~\ref{fig:iono_rms}(c)) and can be represented by the best-fit power law $\propto 1/f^{1.62}$. It should be noted that these low TEC values are derived from the current best estimates of the GPS-TEC errors \citep{hernandez09}.

\item Hence, the simulated spectra, derived every 1 second, is given by $T_{obs}(\nu,TEC_{observed}(t);\Theta_0,\Phi_0)=T_{Ant}^{iono}(\nu,TEC_{observed}(t);\Theta_0,\Phi_0) + T_n$. 

\item The residual spectra is given by $T_{obs}(\nu,TEC_{observed}(t);\Theta_0,\Phi_0)-T^{iono}_{Ant}(\nu,TEC_{model}(t),\Theta_0,\Phi_0)$. RMS of these residual spectra is calculated over 0.5 MHz channel-width and plotted in Figure~\ref{fig:iono_rms}(d).
\end{itemize}

Hence, the uncertainties in the GPS-TEC values still contribute to a residual ionospheric effect in the ionosphere-calibrated spectrum. Figure~\ref{fig:iono_rms}(d) shows the RMS variations due to these inaccuracies in the GPS-TEC measurements near the location of three turning points (B,C and D). It is evident that within the accuracies of the current GPS-TEC measurements it is not possible to reach the desired noise floor of $\sim 1$~mK \citep{burns12} to detect the 3 turning points (Figure~\ref{fig:sig}).   

Although it is not possible to calibrate the ionosphere with the GPS-TEC measurements given their current accuracies, we can assume that with the advancement of GPS technology and ionospheric modeling, uncertainties in the GPS-derived TEC values will decline. For our final simulations, we have assumed that future GPS-TEC measurements will have uncertainties of $\sim 1\%$ of the TEC values measured (i.e. $\sim 0.03$~TECU). In order to examine the effect of this improved accuracy in GPS-TEC measurements, we have performed another simulation over 1000 hours' total integration similar to that in Figure~\ref{fig:iono_rms}(d) but with a different value of $\Delta TEC (t)$. The inaccuracy in the knowledge of TEC measurement or $\Delta TEC (t)$ is now chosen every 1 second from a ``1/f'' process whose power spectrum is plotted in Figure~\ref{fig:iono_rms}(e). Here, the inaccuracy in the TEC measurement is about $1\%$ of that in Figure~\ref{fig:iono_rms}(a). The power spectrum in Figure~\ref{fig:iono_rms}(e) can be represented by the best-fit power law $\propto 1/f^{1.52}$. Figure~\ref{fig:iono_rms}(f) shows the RMS variations due to these inaccuracies in the GPS-TEC measurements near the location of the three turning points `B', `C' and `D'. It is evident that even with the potentially improved accuracy of future GPS-TEC measurements it is still not possible to reach the desired noise floor to detect the three turning points (Figure~\ref{fig:sig}). It should be noted that the frequency locations of the turning points and their magnitudes are highly model-dependent predictions. If turning point D occurs at a lower redshift (or higher frequency, $\gtrsim 100$~MHz), as predicted in \citet{furlanetto06b}, \citet{pritchard08} and \citet{mesinger13}, it may still be possible to detect it from the ground. The effects are more severe for turning points B and C. Hence, we conclude that due to these ionospheric issues, the best chance to detect these two turning points will be from above the Earth's atmosphere \citep{burns12}.  

Independent information about the ionospheric phase and amplitude can be obtained from the radio interferometric observations \citep{bernardi14}. However, it has still to be demonstrated how the information gathered from a radio interferometer can be used to calibrate the ionospheric corruption for a total power experiment. Current state-of-the-art ionospheric calibration has not been able to achieve higher than 1000:1 dynamic range e.g. LOFAR LBA observations at 62 MHz (\citet{vanweeren14}), VLSS 74 MHz all-sky survey (\citet{lane12}). So it will be extremely challenging to use radio interferometers to calibrate the ionosphere in order to extract the faint cosmological 21cm signal with a precision of 1 parts per million.

\section{Conclusion}
\label{sec:conclusion}

In this paper, we have introduced the effects of the dynamic ionosphere -- refraction, absorption and emission -- that affects any trans-ionospheric radio signal. We have also demonstrated the effect of this combined ionospheric contamination on the ground-based global 21 cm signal detection from the Epoch of Reionization and the Cosmic Dawn. Previously, \citet{harish14} showed the effect of ionospheric refraction and absorption on the global 21 cm experiments. This study was based on a static ionosphere and did not include any ionospheric variability. Here, for the first time, we have considered the ionospheric variability and demonstrated its effect on the detection of the global 21 cm signal. 

Due to ionospheric refraction, all sources in the field-of-view appear to move toward the zenith (location of maximum directivity of the antenna). This will result in a further increase in the total power of the radiometer \citep{harish14}. In this paper, we have not explicitly modeled this effect. However, it is evident that inclusion of this effect will only increase the excess sky temperature due to ionospheric refraction (as modeled in this paper) and further deteriorate the prospect of any ground-based detection of the global 21 cm signal.

The variability in the ionospheric TEC was initially derived from the typical night-time conditions at Green Bank, WV, USA (Figures~\ref{fig:TEC_var_hi}(a) and \ref{fig:TEC_var_lo}(a)). The combined effect of ionospheric refraction, absorption and emission creates additional foregrounds which introduces time-dependent spectral features in the residual spectra (Figures~\ref{fig:TEC_var_hi}(b) and \ref{fig:TEC_var_lo}(b)) due to change in the ionospheric TEC values with time. The structure of this additional foreground is a major obstacle in detecting the faint global 21 cm signal, which also shows similar spectral features but at much lower level. We have compared the results from our simulation and modelling with the observed effects of the ionosphere from EDGES data \citep{rogers15}. Our results are consistent with their derived values for the opacity and temperature of the ionosphere.

We have considered the effects of uncertainties in GPS-TEC measurements which will influence the accuracy of any ionospheric calibration scheme. We considered two scenarios, based on the current uncertainties in the GPS-TEC measurements at the $10 \%$ level, and future improvements in the GPS-TEC measurements up to the $1\%$ level. The results in Figures~\ref{fig:iono_rms}(d) and (e) show that with the current and improved accuracies it is not possible to detect any of the three turning points in the model 21 cm signal (Figure~\ref{fig:sig}). However, with the improved accuracies in the GPS-TEC measurements it may be possible to detect turning point 'D' if it occurs at a higher frequency, $\gtrsim 100$~MHz (or lower redshifts). In addition, we have also discussed in appendix~\ref{sec:pink} the strong requirements on any other idealistic ionospheric calibration in order to detect the faint 21cm signal using ground-based observations.

In the simulations, performed in Section~\ref{sec:errors}, we have used a 1 second cadence to denote time interval for ionospheric calibration. It should be noted here that this is an optimistic assumption. In practice, the signal-to-noise over 1 second interval may not be sufficient to even get an accurate ionospheric calibration. Hence, the results shown in Figure 7 are still highly optimistic predictions and in practice the required accuracies on the ionospheric calibration should be higher than mentioned in Section~\ref{sec:errors}.

In the previous section, we have only considered the uncertainties in the GPS-TEC measurements. The variation in the electron temperature ($T_e$) is also another major source of error. $T_e$ is not measured by the GPS observations and requires separate experiments like HF back-scatter radar \citep{schunk78}. It can also be derived from ionospheric models like IRI, NeQUICK, etc \citep{attila96a, attila96b, bilitza03, bilitza15}. The ionospheric models and other experiments have separate sources of errors. It is beyond the scope of this paper to quantify all those uncertainties. However, we can conclude that the total uncertainties in the ionospheric parameters will certainly increase when GPS-TEC measurements are combined with these models and experiments. Hence, the uncertainties in the ionospheric measurements, considered in this paper, still represents the best possible scenario. Moreover, the relative contributions of the electron densities in the D-layer and F-layer to the total column density of electrons in the GPS-TEC measurements is also a model dependent result. In our simulations, we have chosen a typical ratio based on the IRI model. But this ratio can change based on specific geo-location and solar activity. There are other experiments like radio-occultation \citep{jakowski04, attila10}, ionospheric sounding, etc. which when combined with the GPS-TEC measurements and ionospheric models can derive the profile of the electron density \citep{attila97}. The sources of error for all these other experiments have to be considered in order to understand the total uncertainties in the measured ionospheric parameters. 
In this paper, we have not included the contribution from the E layer of the ionosphere. It is expected that the additional consideration of the E layer will only further deteriorate the prospect of any global signal detection from the ground.

Here, we have confirmed the existence of a flicker noise property in the dynamical fluctuations of the ionospheric electron density. These fluctuations directly influence the excess sky noise introduced into the ground based observations at these low radio frequencies. Thus, the additional ionospheric noise in a global 21cm signal data has a flicker noise component which will not integrate down with longer observations. Any attempt to calibrate this noise is subjected to the accuracy in the measurement of the ionospheric parameters. 

Now, we consider a case where only night-time data (4 hours after the local mid-night) is used for analysis. Also, the ionosphere data is assumed to be uncorrelated at time-scales beyond few hours which makes the data from successive nights totally uncorrelated. Hence, flicker noise is only assumed to be dominant within each 4 hours data-sets but not between successive nights. Under the above assumptions, we can use Figure~\ref{fig:iono_rms}b to estimate the total RMS noise after 4 hours of integration to be about 10K near turning point B. Since the ionosphere is assumed to be uncorrelated  from one night to the next, we obtain a 4 hour data-set once each night. Hence, the data are indeed statistically independent night to night. So, theoretically, the central limit theorem states that these "samples" should integrate down.  However, the mean values of each sample (one per day) form the ensemble and it is the ensemble average that integrates down as $\propto 1/\sqrt{N_{days}}$.  So, to achieve 1 mK sensitivity (required sensitivity to detect turning point B \citep{burns12}) would require $10^8$~days or $2.7 \times 10^5$~years, which is quite impractical. Even under the best of circumstances due to improved accuracy in ionospheric calibration, we can take the case for Figure~\ref{fig:iono_rms}f. Here, the total RMS noise after 4 hours of integration is about 1K near turning point B. In this case, we need $10^6$~days or $2,740$~years to reach the requried accuracy of $\sim 1$~mK. Even if we relax the required sensitivity to 10 mK, the required number of years will be around 27.4 years. Even under these idealized situations, our prediction shows that it is quite challenging if not impossible to detect these faint turning points in global 21cm signal from the ground.

In this paper, we found that ionospheric calibration is critical to perform any global 21 cm signal detection from the ground. Under the assumptions of: (i) improved accuracies in future GPS-TEC measurements, and (ii) occurrence of turning point `D' at a higher frequency ($\gtrsim 100$~MHz), the ionospheric effects may be overcome to yield a detection of the turning point D from the ground. However, the ionospheric effects will be a significant obstacle in the detection of the other two turning points (B and C). So, we conclude that space-based observations above the Earth's atmosphere is best suited to detect the crucial turning points B and C below $100$~MHz.

{\bf Acknowledgements:}
AD would like to thank Dayton Jones, Harish Vedantham and Alan Rogers for useful discussions. AD would like to thank Jordan Mirocha for providing the model 21 cm signal. We acknowledge funding from NASA Ames Director's Office in support of the DARE Mission Concept. AD and JOB also acknowledge support from NASA Lunar Science Institute (via Cooperative Agreement NNA09DB30A). GH acknowledges funding from the the People Programme (Marie Curie Actions) of the European Union's Seventh Framework Programme (FP7/2007-2013) under REA grant agreement no. 327999. Part of this research was carried out at the Jet Propulsion Laboratory, California Institute of Technology, under a contract with the National Aeronautics and Space Administration.

\appendix{
\section{Overview of Flicker Noise}
\label{sec:pink}

The statistics of random processes within a dynamical system will affect the accuracy of a measurement and place operational constraints on the nature of the calibration process. Thermal noise sources such as those encountered in astronomy or within the resistances of circuits exhibit the familiar Gaussian statistics having zero mean and non-zero variance (see Figure~\ref{fig:wgn}(a)), leading to a non-zero available power. They are time invariant or stationary random processes, allowing short bursts of non-contiguous power data to be averaged together to improve upon an estimate of its mean value, the reduction in the error follows the well-known standard error model in terms of the radiometer equation:
\begin{equation}
\sigma (\nu) = \frac{T_{sys}(\nu)}{\sqrt{\delta \nu * \delta t}}
\end{equation}
where the symbols have the same meaning as in equation~\ref{eq:therm_noise}.

In theory, only one calibration is required and the scan time can be set to that required by the precision of the measurement, $\delta t = t_{total}$ where $t_{total}$. However, radiometric measurements of the sky obtained by an antenna located on the surface of the Earth will contain fluctuations imposed by the variability of the ionosphere, as described in Section~\ref{sec:effects}, which perturb the Gaussian statistics of the signals through a multiplicative process (see equation~\ref{eq:rte}).  While we are accustom to believing that the Central Limit Theorem will prevail, this assumption is restricted to sums of random variables having finite variances. In contrast, random variables with power law tail distributions, such as those with $1/f^\alpha$ (where $0 \lesssim \alpha \lesssim 2.5$; see Section~\ref{sec:typical}), have infinite variance and will tend to an alpha-stable distribution with a time dependent (non-stationary) mean. The time series and dynamical power spectra for these two cases are shown in Figure~\ref{fig:wgn}. The sky measurement will therefore contain a composite of these two sources of noise: Gaussian white noise and the flicker ``$1/f$'' noise.  Precise, periodic calibrations of the ionosphere are required to remove the flicker component,  yielding a residual that is described only by Gaussian statistics and will thus follow the standard error process.

This periodic calibration, also known as baseline subtraction, will bound the variance of the flicker process only if the residual error after calibration has Gaussian statistics.  It can be shown that the variance per calibration period of a flicker noise process is given by \citep{wilmshurst90}:
\begin{equation}
\sigma_{1/f} \propto A*\ln(t_{scan}/t_{res})
\label{eq:flicker}
\end{equation}   
where $A$ is the amplitude of the power spectrum of a flicker noise, $t_{scan}$ is the time between calibrations and $t_{res}$ is the time per data burst \citep{wilmshurst90}. If an idealized calibration is performed for each data burst such that $t_{scan} = t_{res}$,  then the flicker noise component is removed completely and no additional noise is added to the measurement. It should be noted here that removal of the flicker noise in this case is only accurate to the level of white noise present in the measurement. Moreover, if it takes some time to acquire the idealized baseline data needed for the calibration such that $t_{scan} \gg t_{res}$, then according to equation~\ref{eq:flicker}, the variance of the data over time $t_{scan}$ is non-zero and will contribute a significant amount of Gaussian noise to the measurement even for this idealized case.  The data after calibration will average down as per the standard error process, but the effective system temperature is higher, resulting in a longer integration time to achieve a desired precision.
\begin{figure*}[t!]
\centering
\epsfig{file=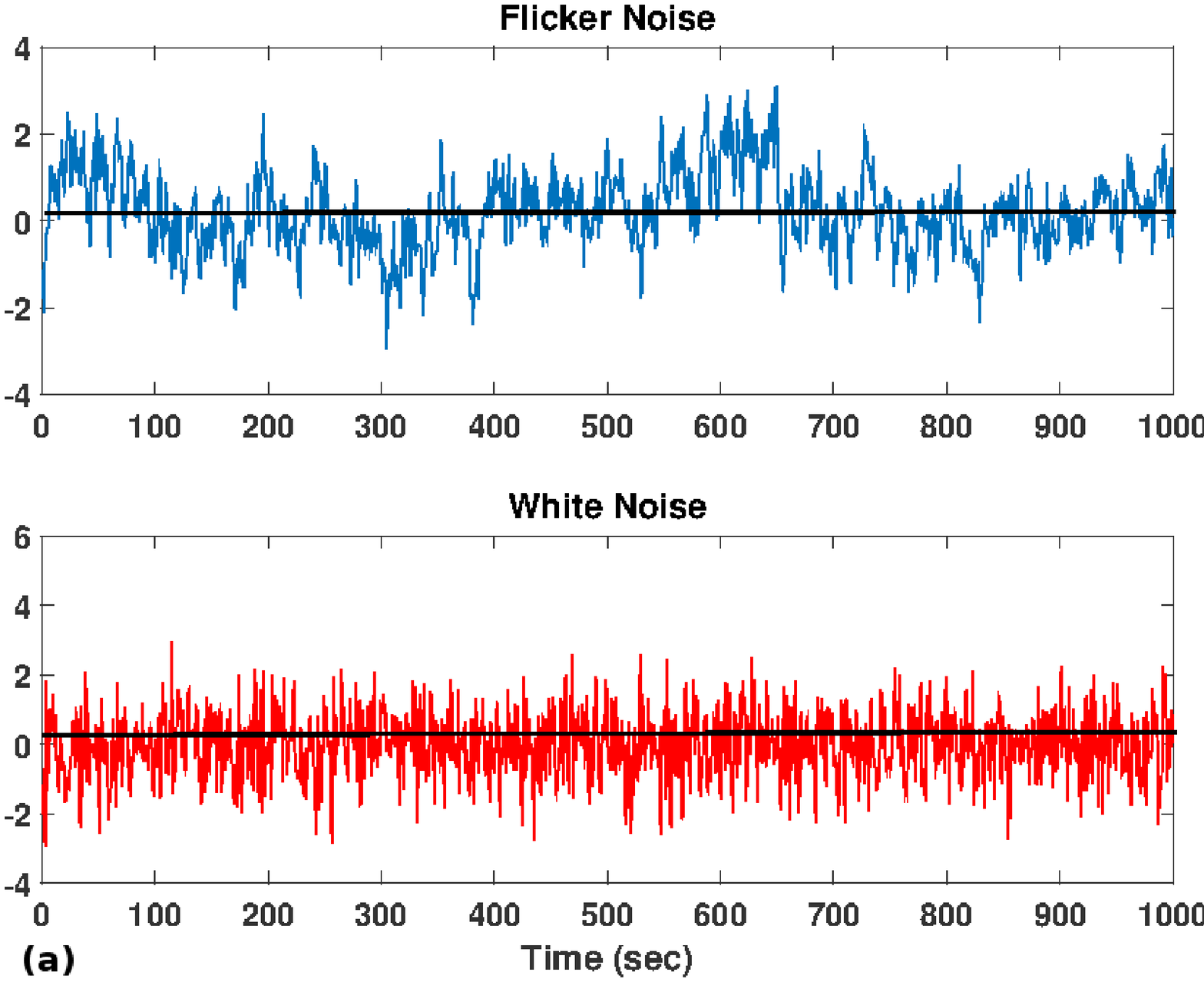,height=2.4truein}
\epsfig{file=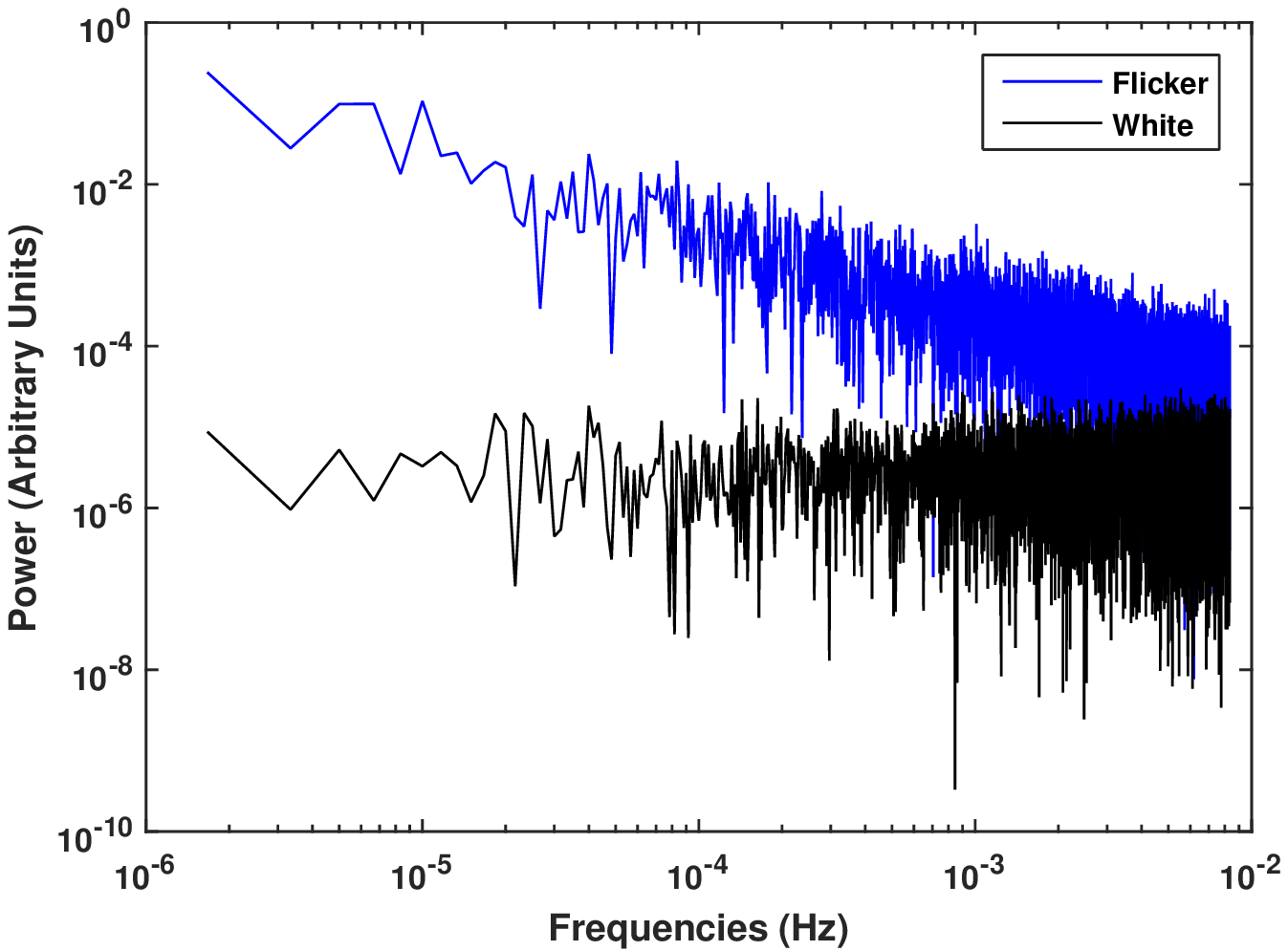,height=2.4truein}
\caption{{\bf (a)} The time series of flicker and white noise. While the mean value for the white noise is fixed, the mean value for the flicker noise varies with time. {\bf (b)} Power spectrum of a flicker noise (in blue) and a white Gaussian noise (in black). It should be mentioned here that the power in the flicker noise has a $1/f$ dependence as expected. On the other hand, the power spectrum of the white noise is flat across dynamical frequency ($f$).}
\label{fig:wgn}
\end{figure*}

Unfortunately, since the ground-based antenna is responding to signals over a rather large region of the sky, an ionospheric calibration will require a precise, rapid measurement of the ionosphere's physical characteristics over this entire sky region during the time $t_{scan}$.  Any residual flicker noise remaining in the data after calibration will appear unbounded (non-stationary) and set a lower limit on the precision that can achieved by the measurement.  Therefore, the variance of the three statistically independent components of the sky measurement (not including the radiometer contribution) is:
\begin{equation}
\sigma_{total}^2= \left(\frac{T_{sky}^2}{\delta \nu * t_{total}} \right) + \left(\frac{T_{FC}^2}{\delta \nu * t_{total}} \right) + T_{FR}^2
\label{eq:final}
\end{equation} 
where $T_{FC}^2=A_{FC}*\ln(t_{scan}/t_{res})$, $T_{FR}^2=A_{FR}*\ln(t_{total}/t_{res})$, $A_{FC}$ is normalized power for the calibrated flicker Gaussian noise and $A_{FR}$  is the normalized power for the residual flicker noise. The first two terms in equation~\ref{eq:final} integrate down over the measurement time, $t_{total}$, which is set by the precision requirements for the science.  The last term will grow in an unbounded manner. 

To meet the Dark Ages science objective, the third term must remain under $1$~mK after the total integration of $t_{total}$ \citep{burns12}. A given ionospheric calibration technique or procedure must clearly demonstrate this level of effectiveness to be viable for Dark Ages science. The models in Figure~\ref{fig:iono_rms} indicate that residual ionospheric flicker noise produce a floor of $\sim 1$~K at 60 MHz, well above that required to observe the turning points. A lunar orbiting spacecraft approach to this measurement will force the second and third terms of equation~\ref{eq:final} to vanish leaving only the Gaussian sky component.

\section{Ionospheric Conditions Across the Globe}
\label{sec:global}
 
The GPS-TEC values also strongly depend on the time of the day (see Section~\ref{sec:measure}), specific location on the Earth and solar activity. Figures~\ref{fig:TEC_var_world} and \ref{fig:TEC_GB} show the typical TEC variation over 5 representative geo-locations with low-latitude ionosphere(Western Australia and South Africa), mid-latitude ionosphere (Netherlands and USA (Green Bank, WV)) and high-latitude ionosphere (Antarctica). It should also be noted that the locations in Western Australia, South Africa and Netherlands are near the sites of current and/or future low-frequency radio telescopes operating above and/or below 100 MHz. These locations are chosen to capture the nature of the variation in the GPS-TEC values across the world: (a) when the solar activity is high in the years 2000 (last Solar Maximum) and 2014 (approaching to the next maximum), (b) when the solar activity is low in the years 2009 and 2010 (last solar minimum). Based on these two figures, we conclude that the night-time GPS-TEC variation at Green Bank, USA over the last solar minimum is similar to any other sites in our sample.
\begin{figure*}[t!]
\centering
\epsfig{file=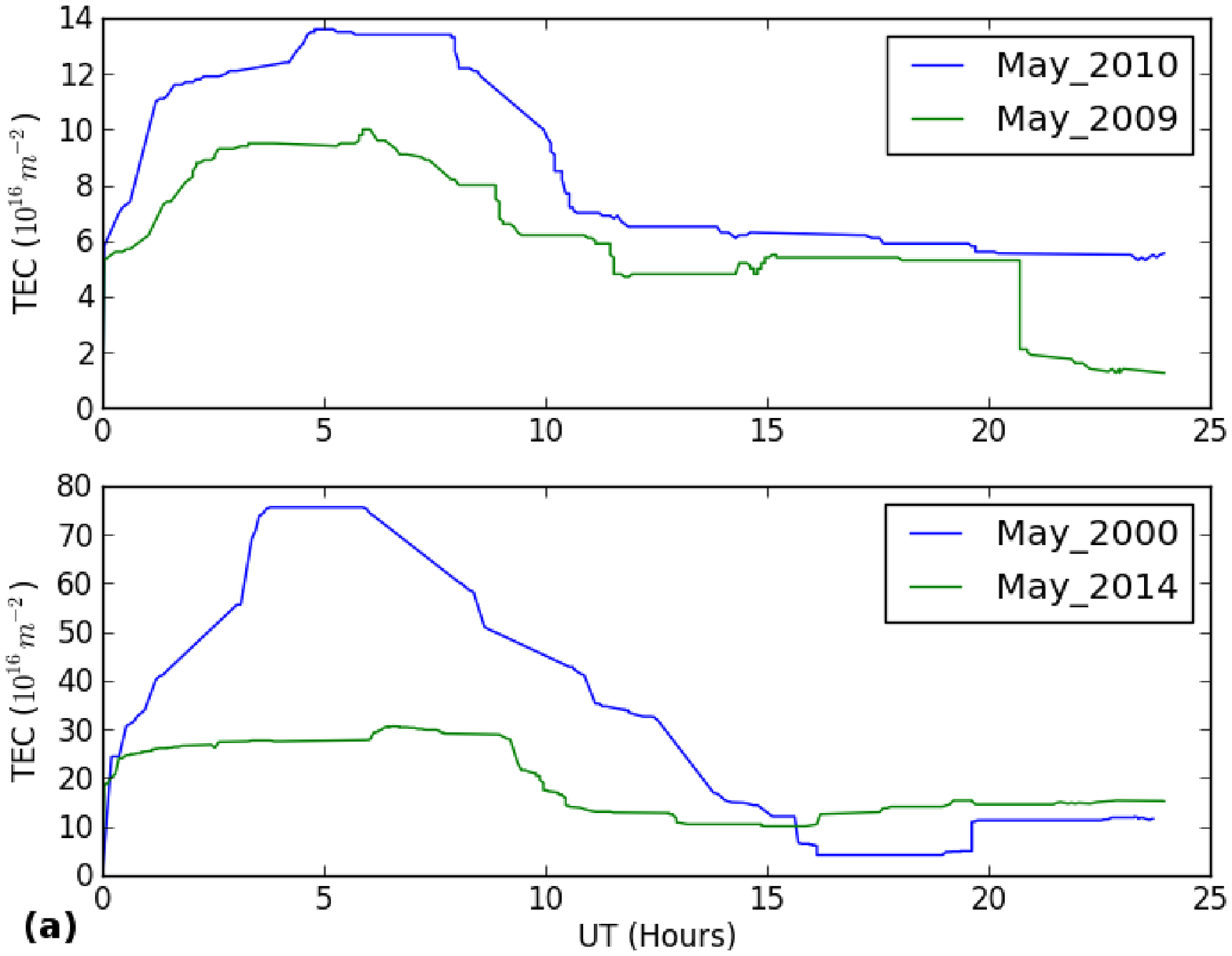,height=2.4truein}
\epsfig{file=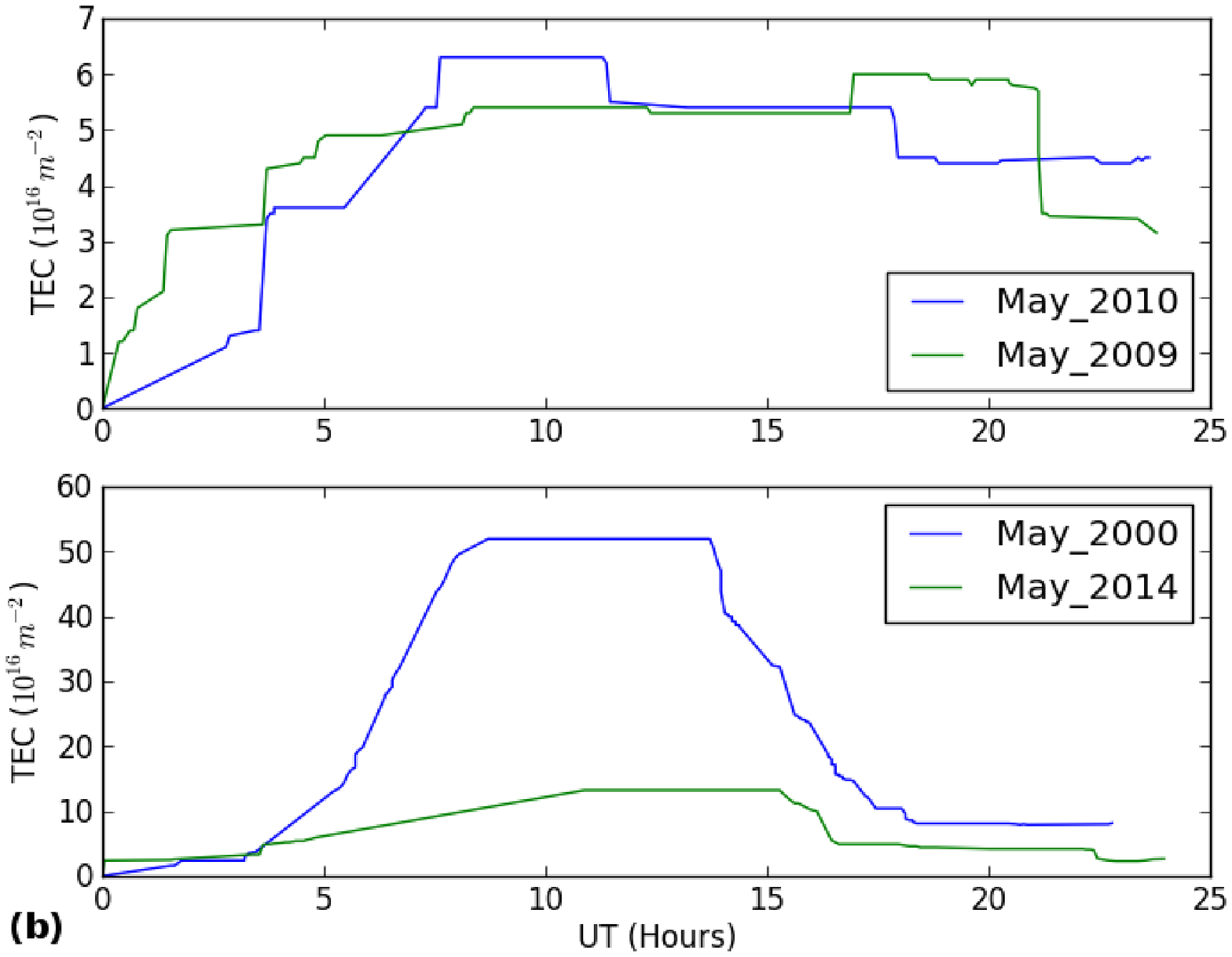,height=2.4truein}
\epsfig{file=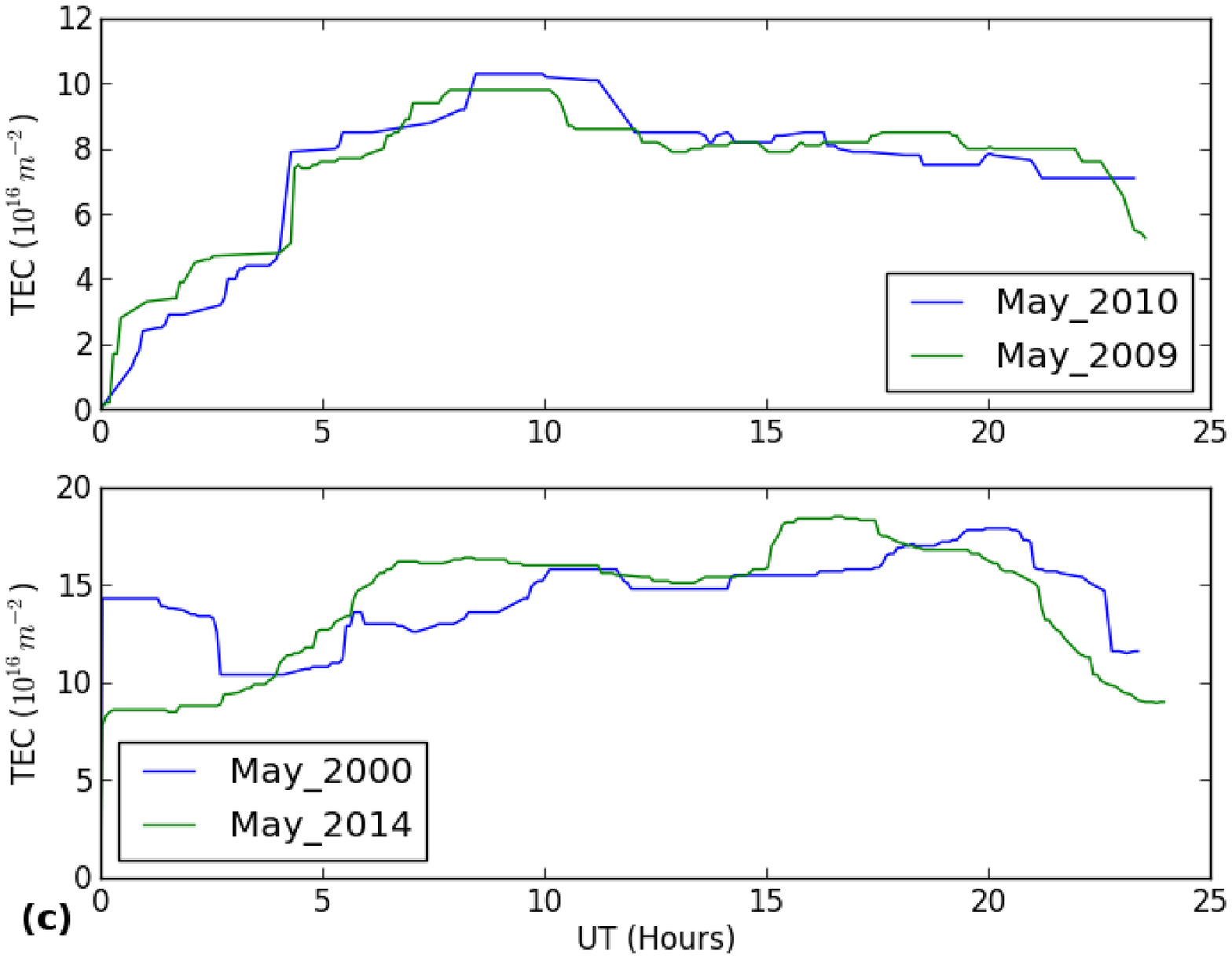,height=2.4truein}
\epsfig{file=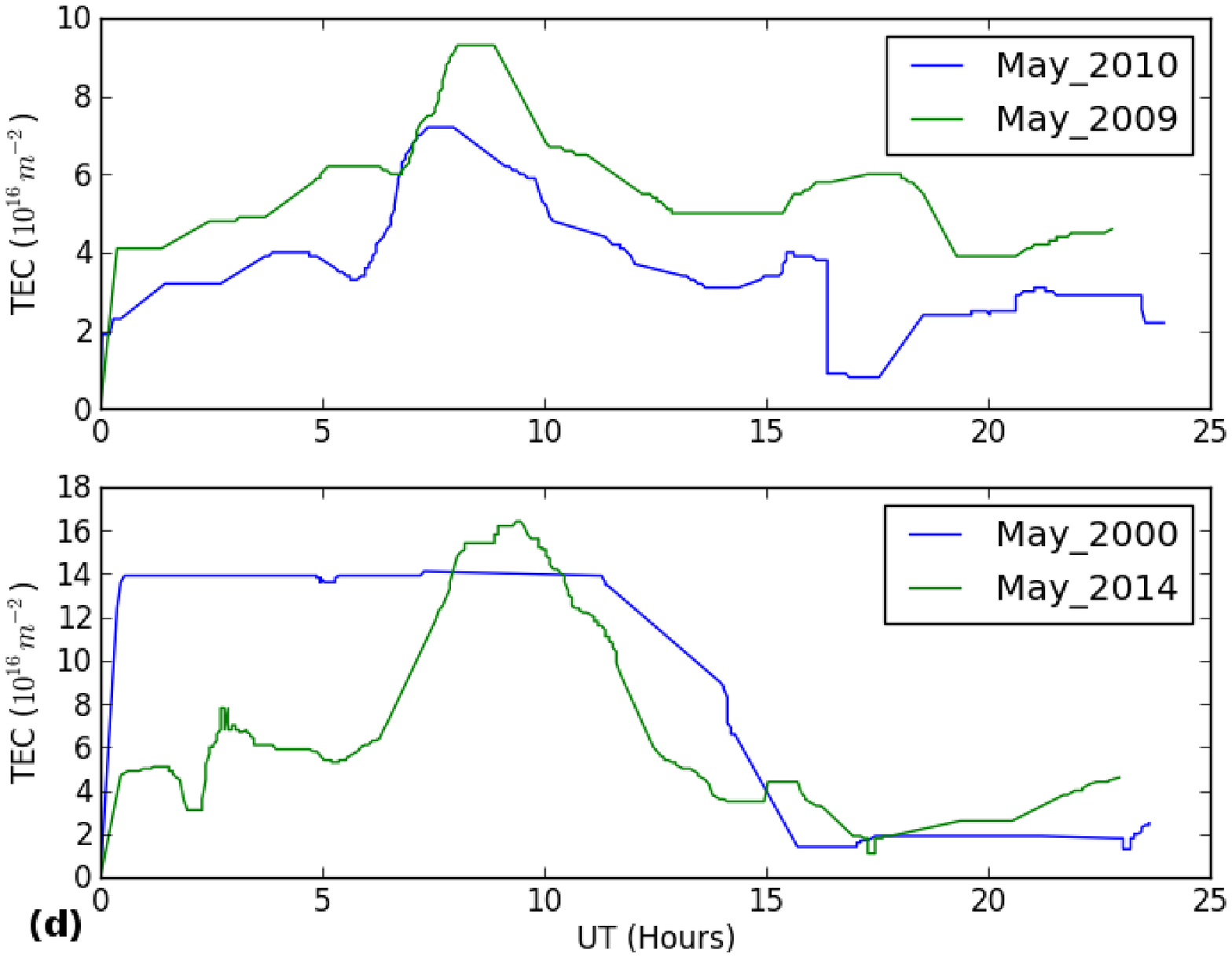,height=2.4truein}
\caption{Variation in the GPS-measured TEC across 4 different sites in the world: {\bf (a)} Australia (Latitude = 26$^o$S and Longitude = 116$^o$E), {\bf (b)} South Africa (Latitude = 31$^o$S and Longitude = 21$^o$E), {\bf (c)} Netherlands (Latitude = 53$^o$N and Longitude = 7$^o$E), and  {\bf (d)} Antarctica (Latitude = 69$^o$S and Longitude = 40$^o$E). The plots include TEC variation near the last Solar Maximum in the year 2000 and near the last Solar Minimum near 2009-2010. All the data for these plots have been obtained from the Madrigal database for the ``World-wide GPS Network'' \citep{rideout06}. It should be noted that the time resolution in available GPS-TEC data varies over different sites. The data over Antarctica is has the lowest time resolution while the data for Netherlands has the highest time resolution.}
\label{fig:TEC_var_world}
\end{figure*}

\begin{figure}[t!]
\centering
\epsfig{file=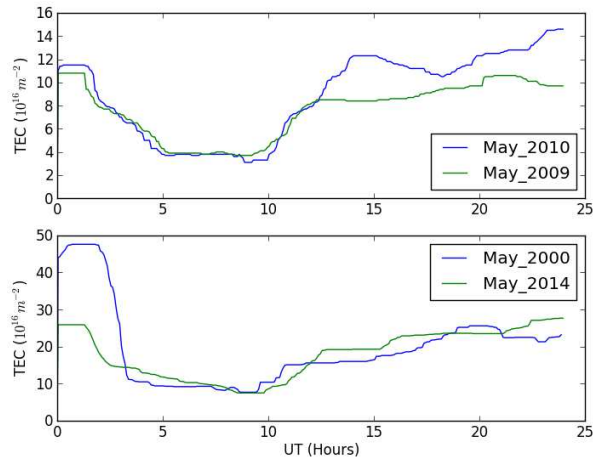,height=2.4truein}
\caption{Variation in the GPS-measured TEC across Green Bank, WV, USA (Latitude= 38$^o$N and Longitude = 80$^o$W). The GPS-TEC data used in these plots have been obtained from the Madrigal database for the ``World-wide GPS Network'' \citep{rideout06}.}
\label{fig:TEC_GB}
\end{figure}
\bibliographystyle{apj}

\end{document}